\documentclass{IEEEtran}
\makeatletter

\usepackage{mathtools,amssymb,yhmath,braket}\interdisplaylinepenalty=2500
	\DeclareMathOperator*\depth{depth}
	\DeclareMathOperator*\bC{bC}

\usepackage[binary-units]{siunitx}
	
	\def\textrightarrow{\rightarrow}

\usepackage{amsthm}
	\newtheorem{thm}{Theorem}
	\newtheorem{lem}[thm]{Lemma}
	
	\newtheorem{prop}[thm]{Proposition}

\usepackage{tikz}
	\definecolor{encolor}{HTML}{BBEECC}
	\definecolor{decolor}{HTML}{DEC0DE}
	\tikzset{
		every picture/.style={
			cap=round,join=round,
			baseline={([yshift=-.5ex]current bounding box)},
		},
		rp/.style={remember picture},o/.style={overlay},
		hl/.style={help lines},
		<^/.style={above left},^/.style={above},^>/.style={above right},
		<</.style={      left}                 ,>>/.style={      right},
		<_/.style={below left},_/.style={below},_>/.style={below right},
	}
	\tikzset{
		bb/.style={/utils/exec=\bbsetupcs,nodes={inner ysep=1,inner xsep=0}},
		G/.pic={\def\coor{coordinate}\path
			(-  \bbwire,   \bbgape)\coor(1L)(   \bbwire,   \bbgape)\coor(1R)
			(-  \bbwire,-  \bbgape)\coor(0L)(   \bbwire,-  \bbgape)\coor(0R);},
		F/.pic={\draw[pic actions]pic{G}(1L)--(1R)(0L)--(0R);},
		E/.pic={\pic[thick]{F};\draw[thick,fill=encolor]
			(-  \bbsize,-  \bbsize)rectangle(   \bbsize,   \bbsize)
			(-  \bbgape,-  \bbgape)       --(   \bbgape,   \bbgape);},
		D/.pic={\pic[thick]{F};\draw[thick,fill=decolor]
			(-  \bbsize,-  \bbsize)rectangle(   \bbsize,   \bbsize)
			(-  \bbgape,   \bbgape)       --(   \bbgape,-  \bbgape)..controls
			( .5\bbgape,-.5\bbgape)    and  ( .5\bbgape, .5\bbgape)..
			(   \bbgape,   \bbgape)       --(-  \bbgape,-  \bbgape);},
	}
	\newdimen\bbsize\bbsize8pt  
	\newdimen\bbwire\bbwire12pt 
	\newdimen\bbgape\bbgape4pt  
	\pgfmathdeclarefunction{xoffset}{1}{\pgfmathparse{
		((2^#1-1)*\bbwire-2*#1\bbgape)/3}}
	\def\bbsetupcs{
		\xdef\bbxmax{0}
		\def\pgfpointxyz##1##2##3{
			\ifnum##2>\bbxmax\xdef\bbxmax{##2}\fi
			\pgfpoint
				{(2\bbwire*##2+xoffset(\bbxmax)-xoffset(\bbxmax+1-##2))*##11}
				{0b##3*2\bbwire}
		}
	}
	\tikzset{
		ut/.style={
			/utils/exec=\utsetupcs,
			nodes={inner sep=1,minimum size=2,scale=2.5^(\utdepth)/3^(\utdepth)},
			baseline=.5\utupper-.5ex
		}
	}
	\def\utdepth{0}
	\newdimen\utupper\utupper8\bbwire
	\newdimen\utmiddle
	\newdimen\utlower\utlower0cm
	\def\utnode$#1$#2$#3#4{
		\utmiddle\dimexpr.5\utupper+.5\utlower\relax
		\draw(\utdepth,\utmiddle)node(\utdepth){$#1$}
			\ifnum\utdepth>0(\the\numexpr\utdepth-1)--(\utdepth)\fi;
		\edef\utdepth{\the\numexpr\utdepth+1}
		{\utlower\utmiddle#3}{\utupper\utmiddle#4}
		\draw[o](\utdepth-.333,\utmiddle)node{$#2$};
	}
	\def\utsetupcs{
		\def\pgfpointxy##1##2{\pgfpoint{2\bbwire*(1.2^(##1)*5-5)}{0pt}}
	}
	\def\TArule#1{\[\tikz\node[rounded corners,draw,align=center]{#1};\]}

\usepackage{pgfplots}
	\pgfplotsset{compat=1.14}
	\usepgfplotslibrary{units}
	\pgfplotsset{
		ylabel style={yshift=-6,o},
		simu/.style={
			ymin=0,xmin=0,
		}
	}

\usepackage[utf8]{inputenc}\let\DUC\DeclareUnicodeCharacter
	\DUC{03B1}\alpha     
	\DUC{03B2}\beta      
	\DUC{03B3}\gamma     
	\DUC{03B4}\delta     
	\DUC{03F5}\epsilon   
	\DUC{03B6}\zeta      
	\DUC{03B7}\eta       
	\DUC{03B8}\theta     
	\DUC{03B9}\iota      
	\DUC{03BA}\kappa     
	\DUC{03BB}\lambda    
	\DUC{03BC}\mu        
	\DUC{03BD}\nu        
	\DUC{03BE}\xi        
	\DUC{03C0}\pi        
	\DUC{03C1}\rho       
	\DUC{03C3}\sigma     
	\DUC{03C4}\tau       
	\DUC{03C5}\upsilon   
	\DUC{03D5}\phi       
	\DUC{03C7}\chi       
	\DUC{03C8}\psi       
	\DUC{03C9}\omega     
	\DUC{0393}\Gamma     
	\DUC{0394}\Delta     
	\DUC{03B5}\varepsilon
	\DUC{0398}\Theta     
	\DUC{03D1}\vartheta  
	\DUC{039B}\Lambda    
	\DUC{039E}\Xi        
	\DUC{03A0}\Pi        
	\DUC{03D6}\varpi     
	\DUC{03F1}\varrho    
	\DUC{03A3}\Sigma     
	\DUC{03C2}\varsigma  
	\DUC{03A5}\Upsilon   
	\DUC{03A6}\Phi       
	\DUC{03C6}\varphi    
	\DUC{03A8}\Psi       
	\DUC{03A9}\Omega     
	
	\DUC{1D53C}{\mathbb E}  
	\DUC{1D53D}{\mathbb F}  
	\DUC{1D540}{\mathbb I}  
	\DUC{2115}{\mathbb N}   
	\DUC{2119}{\mathbb P}   
	\DUC{211A}{\mathbb Q}   
	\DUC{211D}{\mathbb R}   
	\DUC{1D49C}{\mathcal A} 
	\DUC{1D49E}{\mathcal C} 
	\DUC{1D49F}{\mathcal D} 
	\DUC{1D4AF}{\mathcal T} 
	\DUC{2113}\ell 
	
	\DUC{00B7}\cdot 
	\DUC{02C6}\hat 
	\DUC{2022}\bullet 
	\DUC{2190}\leftarrow 
	\DUC{2192}\to 
	\DUC{2208}\in 
	\DUC{2211}\sum 
	\DUC{221A}\sqrt 
	\DUC{221E}\infty 
	\DUC{2227}\wedge 
	\DUC{2228}\vee 
	\DUC{222A}\cup 
	\DUC{2248}\approx 
	\DUC{2254}\coloneqq 
	\DUC{2264}\le 
	\DUC{2265}\ge 
	\DUC{2282}\subset 
	\DUC{27F5}\longleftarrow 
	\DUC{27F6}\longrightarrow 
	
	\DUC{FF08}{\left(} 
	\DUC{FF09}{\right)} 
	\DUC{FF3B}{\left[} 
	\DUC{FF3D}{\right]} 
	\DUC{FF5B}{\left\{} 
	\DUC{FF5D}{\right\}} 
	\DUC{FF5C}{\middle|} 
	
	\DUC{266D}\smashflat 
	\DUC{266F}\smashsharp 
	\def\smashflat{{\smash\flat}}
	\def\smashsharp{{\smash\sharp}}

\usepackage[ampersand]{easylist}

\usepackage{soul}

\usepackage[pdfusetitle]{hyperref}

	\def\G(#1){pic(#1){G}}
	\def\F(#1){pic(#1){F}}
	\def\E(#1){pic(#1){E}}
	\def\D(#1){pic(#1){D}}
	\def\latin{\emph}
	\def\caution{}
	\def\[{\begin{equation}}
	\def\]{\end{equation}}
	\def\Capacity{\textnormal{Capacity}}
	\def\Bha{Bhattacharyya}
	\def\Arikan{Ar\i kan}
	\def\Ari{\text{Ar\i}}
	\def\loglogε{\log\lvert\log\varepsilon\rvert}

\begin{document}
	\title {Log-logarithmic Time Pruned Polar Coding \\
	        on Binary Erasure Channels}
	\author{Hsin-Po Wang and Iwan Duursma \\
	        University of Illinois at Urbana--Champaign \\
	        \{hpwang2, duursma\}@illinois.edu}

\maketitle

\begin{abstract}
	%
	A pruned variant of polar coding is reinvented
	for all binary erasure channels.
	For small $ε>0$, we construct codes with
	block length $ε^{-5}$, code rate $\Capacity-ε$, error probability $ε$,
	and encoding and decoding time complexity $O(N\loglogε)$ per block,
	equivalently $O(\loglogε)$ per information bit
	(Propositions \ref{prop:N} to~\ref{prop:R}).
	
	This result also follows if one
	applies systematic polar coding \cite{Arikan11}
	with simplified successive cancelation decoding \cite{AYK11},
	and then analyzes the performance using \cite{GX13} or \cite{MHU16}.
\end{abstract}


\section{Introduction}

\IEEEPARstart{I}{n the}
	theory of two-terminal error correcting codes,
	four of the most important parameters of block codes are
	block length $N$, code rate $R$, error probability $P$,
	and per-bit time complexity $\bC$.
	
	\def\textrightarrow{\to}
	For instance, Shannon
	proves $(R,P)→(\Capacity,0)$ by paying the price $N→∞$ and exponential $\bC$.
	In the moderate deviations regime, $(N,R,P)→(∞,\Capacity,0)$ parametrically
	by paying exponential $\bC$ \cite{AW10,PV10,AW14,Arikan15,HT15}.
	LDPC codes and friends
	achieve acceptable $(N,R,P,\bC)$-tuples for practical use,
	but $N,P,\bC$ are difficult to parameterize
	\cite{KRU13}.
	RA codes and friends
	enjoy bounded $\bC$ but $N,P$ do not parametrize
	\cite{PSU04,PS05}.
	
	Polar codes have all four parameters parameterized.
	For instance, Lemma~\ref{lem:md} implies
	\begin{align}
		&{} (N,R,P,\bC) \\
		&= （N,\Capacity-N^{-1/5},2^{-N^{1/24}},O(\log N)）.
	\end{align}
	
	We provide a pruned variant of polar codes parametrized by
	\begin{align}
		&{} (N,R,P,\bC) \\
		&= （N,\Capacity-N^{-1/5},N^{-1/5},O(\log\log N)） \\
		&= （ε^{-5},\Capacity-ε,ε,O(\loglogε)）
	\end{align}
	over arbitrary binary erasure channels.
	That is, the per-bit complexity is log-logarithmic 
	in $N$, in $P$, and in $\Capacity-R$.
	This justifies the title.
	
	Section~\ref{sec:prelim} introduces
	\Arikan's idea of channel polarization and our generalization.
	Section~\ref{sec:main} states and proves the main result.
	Section~\ref{sec:connect} connects our work with others'.

\section{Preliminary} \label{sec:prelim}

\subsection{Channel polarization}

	Channel polarization \cite{Arikan09}
	is a method to synthesize some channels to form
	some extremely-unreliable channels and some extremely-reliable channels.
	The user then can transmit uncoded messages through extremely-reliable ones
	while transmitting predictable symbols through extremely-unreliable ones.
	
	We summarize channel polarization as follows.
	Say we are going to communicate over this binary erasure channel
	\[\tikz[bb]{\draw[o]
		(-,1,0)\G(E)(+,1,0)\G(D);
		\draw(E1R)--node[above]{$W$}(D1L)
	}.\]
	We have two magic devices
	\[\tikz[bb]\pic{E};\]
	and
	\[\tikz[bb]\pic{D};\]
	such that if we wire two i.i.d.\ instances of $W$ as follows
	\[\tikz[bb]\draw
		(-,1,0)\E(E)(+,1,0)\D(D)
		(E1R)--node[^>]{$W$}(D1L)
		(E0R)--node[<_]{$W$}(D0L)
		(E1L)node[<<]{$A$}(D1R)node[>>]{$B$}
		(E0L)node[<<]{$C$}(D0R)node[>>]{$D$}
	;, \label{eq:cir2}\]
	then pin $A$ to pin $B$ forms a less reliable synthetic channel $W^♭$,
	while pin $C$ to pin $D$ forms a more reliable synthetic channel $W^♯$.
	Graphically, Formula~(\ref{eq:cir2}) is equivalent to
	\[\tikz[bb]\draw
		(E1L)--node[^>]{$W^♭$}(D1R)
		(E0L)--node[<_]{$W^♯$}(D0R)
	;.\]
	
	Formula~(\ref{eq:cir2}) being the base step,
	the next step is to duplicate Formula~(\ref{eq:cir2}) and wire them as
	\[\tikz[bb]{\draw[o]
		(-,2,1)\G(1)(-,1,1)\E(2)(+,1,1)\D(3)(+,2,1)\G(4)
		(-,2,0)\G(5)(-,1,0)\E(6)(+,1,0)\D(7)(+,2,0)\G(8)
	;\draw
		(11R)--(21L)(21R)--node[^>]{$W$}(31L)(31R)--(41L)
		(51R)--(20L)(20R)--node[<_]{$W$}(30L)(30R)--(81L)
		(10R)--(61L)(61R)--node[^>]{$W$}(71L)(71R)--(40L)
		(50R)--(60L)(60R)--node[<_]{$W$}(70L)(70R)--(80L)
	;}, \label{eq:xcir2x}\]
	which is equivalent to four synthetic channels as
	\[\tikz[bb]{\draw
		(11R)--(21L)--node[^>]{$W^♭$}(31R)--(41L)
		(51R)--(20L)--node[<_]{$W^♯$}(30R)--(81L)
		(10R)--(61L)--node[^>]{$W^♭$}(71R)--(40L)
		(50R)--(60L)--node[<_]{$W^♯$}(70R)--(80L)
	;}\]
	or simply
	\[\tikz[bb]\draw
		(11R)--node[^>]{$W^♭$}(41L)
		(10R)--node[<_]{$W^♭$}(40L)
		(51R)--node[^>]{$W^♯$}(81L)
		(50R)--node[<_]{$W^♯$}(80L)
	;.\]
	Further wire Formula~(\ref{eq:xcir2x}) as
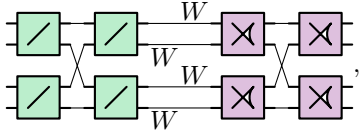
	\[\tikz[bb]\draw
		(-,2,1)\E(1)(-,1,1)\E(2)(+,1,1)\D(3)(+,2,1)\D(4)
		(-,2,0)\E(5)(-,1,0)\E(6)(+,1,0)\D(7)(+,2,0)\D(8)
		(11R)--(21L)(21R)--node[^>]{$W$}(31L)(31R)--(41L)
		(10R)--(61L)(20R)--node[<_]{$W$}(30L)(30R)--(81L)
		(51R)--(20L)(61R)--node[^>]{$W$}(71L)(71R)--(40L)
		(50R)--(60L)(60R)--node[<_]{$W$}(70L)(70R)--(80L)
	;, \label{eq:cir4}\]
	which is equivalent to
	
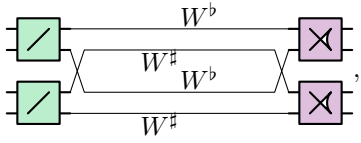
\[\tikz[bb]\draw
		(-,2,1)\E(1)(+,2,1)\D(4)
		(-,2,0)\E(5)(+,2,0)\D(8)
		(11R)--(21L)--node[^>]{$W^♭$}(31R)--(41L)
		(51R)--(20L)--node[<_]{$W^♯$}(30R)--(81L)
		(10R)--(61L)--node[^>]{$W^♭$}(71R)--(40L)
		(50R)--(60L)--node[<_]{$W^♯$}(70R)--(80L)
	;,\]
	to
	
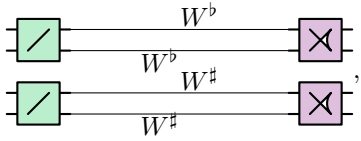
\[\tikz[bb]\draw
		(-,2,1)\E(1)(+,2,1)\D(4)
		(-,2,0)\E(5)(+,2,0)\D(8)
		(11R)--node[^>]{$W^♭$}(41L)
		(10R)--node[<_]{$W^♭$}(40L)
		(51R)--node[^>]{$W^♯$}(81L)
		(50R)--node[<_]{$W^♯$}(80L)
	;,\]
	and to
	
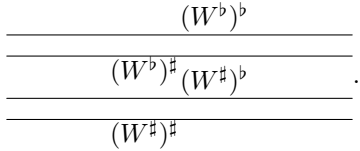
\[\tikz[bb]\draw
		(11L)--node[^>]{$(W^♭)^♭$}(41R)
		(10L)--node[<_]{$(W^♭)^♯$}(40R)
		(51L)--node[^>]{$(W^♯)^♭$}(81R)
		(50L)--node[<_]{$(W^♯)^♯$}(80R)
	;.\]
	Here $(W^♭)^♭$ is a synthetic channel less reliable than $W^♭$;
	synthetic channel $(W^♭)^♯$ is  more reliable than $W^♭$;
	synthetic channel $(W^♯)^♭$ is less reliable than $W^♯$; and
	synthetic channel $(W^♯)^♯$ is more reliable than $W^♯$.
	
	After Formula~(\ref{eq:cir4}), the next, larger construction is
	two copies of Formula~(\ref{eq:cir4}) plus four more pairs of magic devices
	
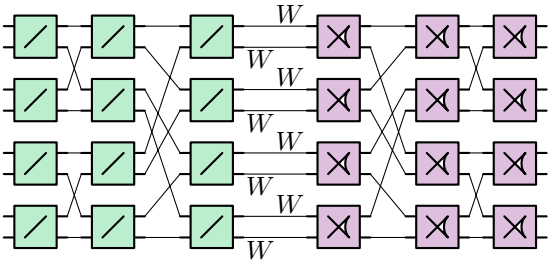
\[\tikz[bb]{\draw
		foreach\J in{00,01,10,11}{
			foreach\l in{3,2,1}{
				(-,\l,\J)\E(\l E\J)(+,\l,\J)\D(\l D\J)}
			(1E\J1R)--node[^>]{$W$}(1D\J1L)
			(1E\J0R)--node[<_]{$W$}(1D\J0L)};
		\def\DRAW#1#2{
			\draw foreach\J in{000,001,010,011,100,101,110,111}{
				(#2E\K R)--(#1E\J L)(#1D\J R)--(#2D\K L)};}
		\def\K{\expandafter\k\J}
		\def\k#1#2#3{#3#1#2}\DRAW12
		\def\k#1#2#3{#1#3#2}\DRAW23
	}. \label{eq:cir8}\]
	It is equivalent to
	\[\tikz[bb]{\draw
		foreach\J in{00,01,10,11}{
			foreach\l in{3,2}{
				(-,\l,\J)\E(\l E\J)(+,\l,\J)\D(\l D\J)}
			(1E\J1L)--node[^>]{$W^♭$}(1D\J1R)
			(1E\J0L)--node[<_]{$W^♯$}(1D\J0R)};
		\def\DRAW#1#2{
			\draw foreach\J in{000,001,010,011,100,101,110,111}{
				(#2E\K R)--(#1E\J L)(#1D\J R)--(#2D\K L)};}
		\def\K{\expandafter\k\J}
		\def\k#1#2#3{#3#1#2}\DRAW12
		\def\k#1#2#3{#1#3#2}\DRAW23
	},\]
	to
	\[\tikz[bb]{\draw
		foreach \J in{00,01,10,11}{
			foreach\l in{3,2}{
				(-,\l,\J)\E(\l E\J)(+,\l,\J)\D(\l D\J)}}
		foreach\J in{0,1}{
			(2E1\J1R)--node[^>]{$W^♭$}(2D1\J1L)
			(2E1\J0R)--node[<_]{$W^♭$}(2D1\J0L)}
		foreach\J in{0,1}{
			(2E0\J1R)--node[^>]{$W^♯$}(2D0\J1L)
			(2E0\J0R)--node[<_]{$W^♯$}(2D0\J0L)};
		\def\DRAW#1#2{
			\draw foreach\J in{000,001,010,011,100,101,110,111}{
				(#2E\K R)--(#1E\J L)(#1D\J R)--(#2D\K L)};}
		\def\K{\expandafter\k\J}
		\def\k#1#2#3{#1#3#2}\DRAW23
	},\]
	to
	\[\tikz[bb]{\draw
		foreach\J in{00,01,10,11}{
			foreach\l in{3}{
				(-,\l,\J)\E(\l E\J)(+,\l,\J)\D(\l D\J)}}
		foreach\J in{0,1}{
			(2E1\J1L)--node[^>]{$(W^♭)^♭$}(2D1\J1R)
			(2E1\J0L)--node[<_]{$(W^♭)^♯$}(2D1\J0R)}
		foreach\J in{0,1}{
			(2E0\J1L)--node[^>]{$(W^♯)^♭$}(2D0\J1R)
			(2E0\J0L)--node[<_]{$(W^♯)^♯$}(2D0\J0R)};
		\def\DRAW#1#2{
			\draw foreach\J in{000,001,010,011,100,101,110,111}{
				(#2E\K R)--(#1E\J L)(#1D\J R)--(#2D\K L)};}
		\def\K{\expandafter\k\J}
		\def\k#1#2#3{#1#3#2}\DRAW23
	},\]
	to
	\[\tikz[bb]{\draw
		foreach\J in{00,01,10,11}{
			foreach\l in{3}{
				(-,\l,\J)\E(\l E\J)(+,\l,\J)\D(\l D\J)}}
		(3E111R)--node[^>]{$(W^♭)^♭$}(3D111L)
		(3E110R)--node[<_]{$(W^♭)^♭$}(3D110L)
		(3E101R)--node[^>]{$(W^♭)^♯$}(3D101L)
		(3E100R)--node[<_]{$(W^♭)^♯$}(3D100L)
		(3E011R)--node[^>]{$(W^♯)^♭$}(3D011L)
		(3E010R)--node[<_]{$(W^♯)^♭$}(3D010L)
		(3E001R)--node[^>]{$(W^♯)^♯$}(3D001L)
		(3E000R)--node[<_]{$(W^♯)^♯$}(3D000L);
	},\]
	and finally to
	\[\tikz[bb]{\draw
		(3E111L)--node[^>]{$((W^♭)^♭)^♭$}(3D111R)
		(3E110L)--node[<_]{$((W^♭)^♭)^♯$}(3D110R)
		(3E101L)--node[^>]{$((W^♭)^♯)^♭$}(3D101R)
		(3E100L)--node[<_]{$((W^♭)^♯)^♯$}(3D100R)
		(3E011L)--node[^>]{$((W^♯)^♭)^♭$}(3D011R)
		(3E010L)--node[<_]{$((W^♯)^♭)^♯$}(3D010R)
		(3E001L)--node[^>]{$((W^♯)^♯)^♭$}(3D001R)
		(3E000L)--node[<_]{$((W^♯)^♯)^♯$}(3D000R);
	}.\]
	Here $((W^♭)^♭)^♭$ is a synthetic channel less reliable than $(W^♭)^♭$; etc.
	
	After Formula~(\ref{eq:cir8}), the next, larger construction
	is going to be two copies of Formula~(\ref{eq:cir8})
	plus one extra layer of magic devices.
	
	The game goes on endlessly.
	\Arikan{} then observes that synthetic channels generated in this way
	tend to be either extremely reliable  or extremely unreliable.
	That is so say, they \emph{polarize}.

\subsection{Channel polarization in Tree Notation}

	Draw
	
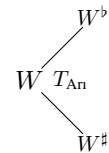
\[\tikz[ut]{
		\utnode$W$T_\Ari${
			\utnode$W^♭$${}{} }{
			\utnode$W^♯$${}{} }
	} \label{eq:tre2}\]
	to capture the fact that Formula~(\ref{eq:cir2})
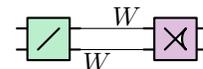
	\begin{equation*}\tikz[bb]\draw
		(-,1,0)\E(E)(+,1,0)\D(D)
		(E1R)--node[^>]{$W$}(D1L)
		(E0R)--node[<_]{$W$}(D0L)
	;\end{equation*}
	transforms two instances of $W$ into a $W^♭$ and a $W^♯$.
	
	Similarly, draw
	
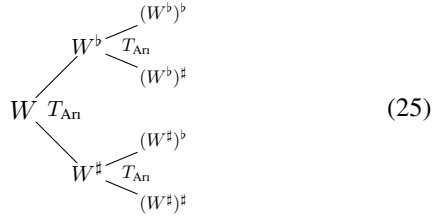
\[\tikz[ut]{
		\utnode$W$T_\Ari${
			\utnode$W^♭$T_\Ari${
				\utnode$(W^♭)^♭$${}{} }{
				\utnode$(W^♭)^♯$${}{} } }{
			\utnode$W^♯$T_\Ari${
				\utnode$(W^♯)^♭$${}{} }{
				\utnode$(W^♯)^♯$${}{} } }
	} \label{eq:tre4}\]
	to capture the fact that Formula~(\ref{eq:cir4})
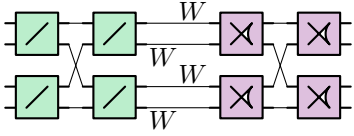
	\begin{equation*}\tikz[bb]\draw
		(-,2,1)\E(1)(-,1,1)\E(2)(+,1,1)\D(3)(+,2,1)\D(4)
		(-,2,0)\E(5)(-,1,0)\E(6)(+,1,0)\D(7)(+,2,0)\D(8)
		(11R)--(21L)(21R)--node[^>]{$W$}(31L)(31R)--(41L)
		(10R)--(61L)(20R)--node[<_]{$W$}(30L)(30R)--(81L)
		(51R)--(20L)(61R)--node[^>]{$W$}(71L)(71R)--(40L)
		(50R)--(60L)(60R)--node[<_]{$W$}(70L)(70R)--(80L)
	;\end{equation*}
	transforms four instances of $W$ into two pairs of $W^♭$ and $W^♯$.
	Two $W^♭$ are then transformed into a $(W^♭)^♭$ and a $(W^♭)^♯$;
	two $W^♯$ are then transformed into a $(W^♯)^♭$ and a $(W^♯)^♯$.
	
	Similarly, draw
	
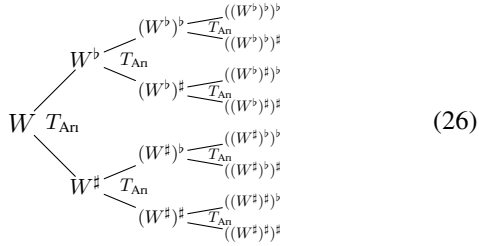
\[\tikz[ut]{
		\utnode$W$T_\Ari${
			\utnode$W^♭$T_\Ari${
				\utnode$(W^♭)^♭$T_\Ari${
					\utnode$((W^♭)^♭)^♭$${}{} }{
					\utnode$((W^♭)^♭)^♯$${}{} } }{
				\utnode$(W^♭)^♯$T_\Ari${
					\utnode$((W^♭)^♯)^♭$${}{} }{
					\utnode$((W^♭)^♯)^♯$${}{} } } }{
			\utnode$W^♯$T_\Ari${
				\utnode$(W^♯)^♭$T_\Ari${
					\utnode$((W^♯)^♭)^♭$${}{} }{
					\utnode$((W^♯)^♭)^♯$${}{} } }{
				\utnode$(W^♯)^♯$T_\Ari${
					\utnode$((W^♯)^♯)^♭$${}{} }{
					\utnode$((W^♯)^♯)^♯$${}{} } } }
	} \label{eq:tre8}\]
	to capture Formula~(\ref{eq:cir8})
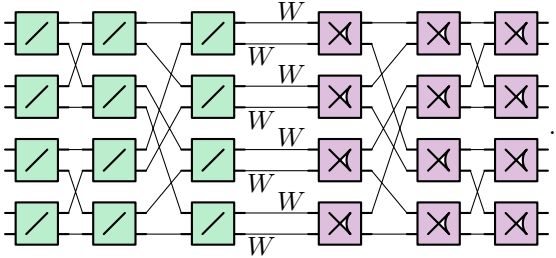
	\begin{equation*}\tikz[bb]{
		\foreach\J in{00,01,10,11}{
			\draw foreach\l in{3,2,1}{
					(-,\l,\J)\E(\l E\J)(+,\l,\J)\D(\l D\J)}
				(1E\J1R)--node[^>]{$W$}(1D\J1L)
				(1E\J0R)--node[<_]{$W$}(1D\J0L);}
		\def\DRAW#1#2{
			\draw foreach\J in{000,001,010,011,100,101,110,111}{
				(#2E\K R)--(#1E\J L)(#1D\J R)--(#2D\K L)};}
		\def\K{\expandafter\k\J}
		\def\k#1#2#3{#3#1#2}\DRAW12
		\def\k#1#2#3{#1#3#2}\DRAW23
	}.\end{equation*}
	That is, eight instances of $W$ are transformed into four pairs of $W^♭,W^♯$,
	into two quadruples of $(W^♭)^♭$, $(W^♭)^♯$, $(W^♯)^♭$, $(W^♯)^♯$,
	and finally into
	$((W^♭)^♭)^♭$, $((W^♭)^♭)^♯$, $((W^♭)^♯)^♭$, $((W^♭)^♯)^♯$,
	$((W^♯)^♭)^♭$, $((W^♯)^♭)^♯$, $((W^♯)^♯)^♭$, $((W^♯)^♯)^♯$.
	
	It is not hard to imagine that the next construction will
	transform sixteen instances of $W$ into ``some intermediate things'',
	and finally into $(((W^♭)^♭)^♭)^♭$ to $(((W^♯)^♯)^♯)^♯$.

\subsection{Generalize to Unbalanced Tree Notation}

	The generalization comes in two perspectives,
	each motivated by an attempt to optimize polar coding.
	
	First perspective: in a tree like Formula~(\ref{eq:tre8}) or a larger tree,
	it could be the case that some synthetic channel, say $(W^♭)^♭$,
	is so bad that applying further transformations sounds useless.
	If so, we may remove children of $(W^♭)^♭$ to get
	
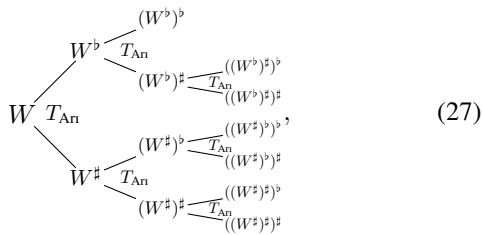
\[\tikz[ut]{
		\utnode$W$T_\Ari${
			\utnode$W^♭$T_\Ari${
				\utnode$(W^♭)^♭$${}{} }{
				\utnode$(W^♭)^♯$T_\Ari${
					\utnode$((W^♭)^♯)^♭$${}{} }{
					\utnode$((W^♭)^♯)^♯$${}{} } } }{
			\utnode$W^♯$T_\Ari${
				\utnode$(W^♯)^♭$T_\Ari${
					\utnode$((W^♯)^♭)^♭$${}{} }{
					\utnode$((W^♯)^♭)^♯$${}{} } }{
				\utnode$(W^♯)^♯$T_\Ari${
					\utnode$((W^♯)^♯)^♭$${}{} }{
					\utnode$((W^♯)^♯)^♯$${}{} } } }
	}, \label{eq:tre7}\]
	which translates into the circuit
	
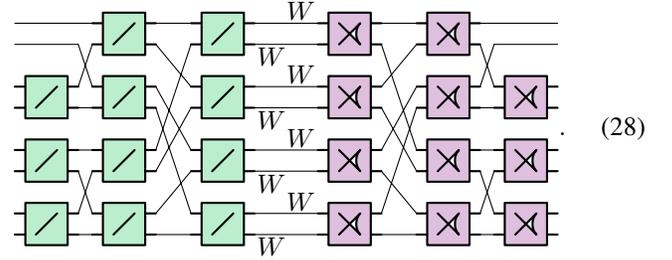
\[\tikz[bb]{
		\foreach\J in{00,01,10,11}{
			\draw foreach\l in{3,2,1}{\ifnum\J\l=113
					(-,\l,\J)\F(\l E\J)(+,\l,\J)\F(\l D\J)\else
					(-,\l,\J)\E(\l E\J)(+,\l,\J)\D(\l D\J)\fi}
				(1E\J1R)--node[^>]{$W$}(1D\J1L)
				(1E\J0R)--node[<_]{$W$}(1D\J0L);}
		\def\DRAW#1#2{
			\draw foreach\J in{000,001,010,011,100,101,110,111}{
				(#2E\K R)--(#1E\J L)(#1D\J R)--(#2D\K L)};}
		\def\K{\expandafter\k\J}
		\def\k#1#2#3{#3#1#2}\DRAW12
		\def\k#1#2#3{#1#3#2}\DRAW23
	}. \label{eq:cir7}\]
	That is, eight instances of $W$ are transformed into four pairs of $W^♭,W^♯$,
	into two quadruples of $(W^♭)^♭$, $(W^♭)^♯$, $(W^♯)^♭$, $(W^♯)^♯$.
	And then, \caution{notice the difference},
	while keeping two $(W^♭)^♭$,
	the other six are transformed into $((W^♭)^♯)^♭$, $((W^♭)^♯)^♯$,
	$((W^♯)^♭)^♭$, $((W^♯)^♭)^♯$, $((W^♯)^♯)^♭$, $((W^♯)^♯)^♯$.
	
	Second perspective: in a tree like Formula~(\ref{eq:tre4}),
	it could be that some synthetic channel, say $(W^♭)^♯$,
	might not be polarized enough,
	i.e. it is neither extremely good nor extremely bad.
	thus we further polarize it by applying an additional $T_\Ari$ as follows:
	
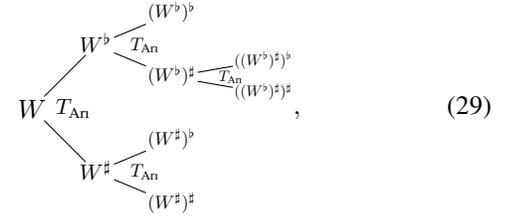
\[\tikz[ut]{
		\utnode$W$T_\Ari${
			\utnode$W^♭$T_\Ari${
				\utnode$(W^♭)^♭$${}{} }{
				\utnode$(W^♭)^♯$T_\Ari${
					\utnode$((W^♭)^♯)^♭$${}{} }{
					\utnode$((W^♭)^♯)^♯$${}{} } } }{
			\utnode$W^♯$T_\Ari${
				\utnode$(W^♯)^♭$${}{} }{
				\utnode$(W^♯)^♯$${}{} } }
	}, \label{eq:tre5}\]
	which translates into the circuit
	
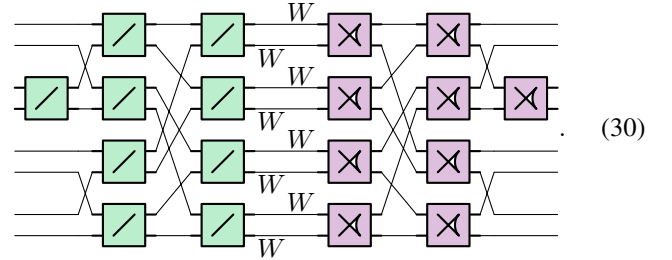
\[\tikz[bb]{
		\foreach\J in{00,01,10,11}{
			\draw foreach\l in{3,2,1}{
				\ifnum\l=3
					\ifnum\J=10
						(-,\l,\J)\E(\l E\J)(+,\l,\J)\D(\l D\J)
					\else
						(-,\l,\J)\F(\l E\J)(+,\l,\J)\F(\l D\J)
					\fi
				\else
					(-,\l,\J)\E(\l E\J)(+,\l,\J)\D(\l D\J)
				\fi}
				(1E\J1R)--node[^>]{$W$}(1D\J1L)
				(1E\J0R)--node[<_]{$W$}(1D\J0L);}
		\def\DRAW#1#2{
			\draw foreach\J in{000,001,010,011,100,101,110,111}{
				(#2E\K R)--(#1E\J L)(#1D\J R)--(#2D\K L)};}
		\def\K{\expandafter\k\J}
		\def\k#1#2#3{#3#1#2}\DRAW12
		\def\k#1#2#3{#1#3#2}\DRAW23
	}. \label{eq:cir5}\]
	That is, eight instances of $W$ are transformed into four pairs of $W^♭,W^♯$,
	into two quadruples of $(W^♭)^♭$, $(W^♭)^♯$, $(W^♯)^♭$, $(W^♯)^♯$,
	and, \caution{notice another differnece},
	only the two $(W^♭)^♯$ are transformed into $((W^♭)^♯)^♭$, $((W^♭)^♯)^♯$.
	
	In general, any rooted, full
	(each vertex has either zero or two children)
	binary tree of channels translates into a circuit of magic devices
	that transforms copies of the root channel to (copies of) leaf channels.
	See Appendix~\ref{app:polish} for more examples.
	
	Denote by $𝒯$ a tree of channels with root channel $W$.
	The number of root channels that $𝒯$ consumes is $2^{\depth(𝒯)}$.
	The number of copies of a leaf channel $w$
	that $𝒯$ synthesizes is $2^{\depth(𝒯)-\depth(w)}$.
	(Convention: the root has depth $0$;
	the depth of a tree is the depth of the deepest leaf;
	and the tree with only one vertex has depth $0$.)

\subsection{\Bha{} Parameter and Processes}

	The \emph{\Bha{} parameter} $Z(W)$ of a channel $W$
	measures the unreliability, the badness, of the channel.
	For binary erasure channels, $Z(W)$ coincides with
	the erasure probability of~$W$.
	The Shannon capacity $I(W)$ of $W$ coincides with the complement $1-Z(W)$.
	
	Recall the processes $K_i$, $Z_i$, and $I_i$ as defined in
	\cite[Section~IV, third paragraph]{Arikan09}.
	We now define their generalizations.
	
	Given a channel tree $𝒯$ with root channel $W$,
	define three discrete-time stochastic processes
	$K_{i∧τ}$, $Z_{i∧τ}$, $I_{i∧τ}$ and a stopping time $τ$ as follows:
	Start from the root channel $K_{0∧τ}≔W$.
	For any $i≥0$, if $K_{i∧τ}$ is a leaf, let $K_{i+1∧τ}$ be $K_{i∧τ}$.
	If, otherwise, $K_{i∧τ}$ has two children,
	choose either child with equal probability as $K_{i+1∧τ}$.
	Let $Z_{i∧τ}$ be $Z(K_{i∧τ})$.
	Let $I_{i∧τ}$ be $I(K_{i∧τ})$.
	Let $K_τ,Z_τ,I_τ$ be the limits.
	Let $τ$ be $\depth(K_τ)$.
	
	By \cite[Proposition~8]{Arikan09}, $I_i$ is a martingale.
	Hence $I_{i∧τ}$ is martingale by \cite[Theorem~5.2.6]{Durrett10}.
	Since $W$ is an erasure channel, $Z_{i∧τ}=1-I_{i∧τ}$ is martingale as well.
	A charming consequence by \cite[Theorem~5.4.1]{Durrett10} is
	\[I(W)=I_0=𝔼[I_τ].\]
	
	For a tree $𝒯$ as in Formula~(\ref{eq:tre5}),
	a possible instance of the process is
	
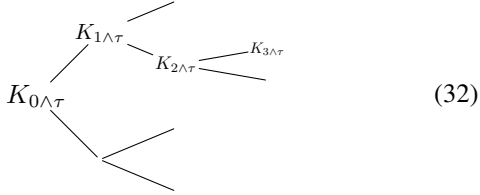
\[\tikz[ut]{
		\utnode$K_{0∧τ}$${
			\utnode$K_{1∧τ}$${
				\utnode$$${}{} }{
				\utnode$K_{2∧τ}$${
					\utnode$K_{3∧τ}$${}{} }{
					\utnode$$${}{} } } }{
			\utnode$$${
				\utnode$$${}{} }{
				\utnode$$${}{} } }
	}\]
	with $K_{3∧τ}=K_{4∧τ}=K_{5∧τ}=\dotsb=K_τ$ and $τ=3$.
	The probability measure of this path is $1/8$.
	For another instance
	
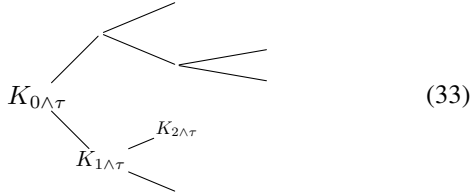
\[\tikz[ut]{
		\utnode$K_{0∧τ}$${
			\utnode$$${
				\utnode$$${}{} }{
				\utnode$$${
					\utnode$$${}{} }{
					\utnode$$${}{} } } }{
			\utnode$K_{1∧τ}$${
				\utnode$K_{2∧τ}$${}{} }{
				\utnode$$${}{} } }
	}\]
	with $K_{2∧τ}=K_{3∧τ}=K_{4∧τ}=\dotsb=K_τ$ and $τ=2$,
	the probability measure is $1/4$.

\subsection{Construct Codes and Communicate}

	In a given tree $𝒯$, non-leaf vertices represent
	channels that are consumed to obtain their children.
	They are not available to users.
	Leaves of $𝒯$, however, represent channels that are available to users.
	
	A person who wants to send messages can
	(a) choose a subset $𝒜$ of leaves,
	(b) transmit uncoded messages through leaf channels in $𝒜$, and
	(c) transmit predictable symbols through the remaining leaf channels.
	
	This tree-leaves pair $(𝒯,𝒜)$ determines a block code.
	A block code has
	block length~$N$, code rate~$R$, error probability~$P$, and time complexity.
	The following is how to read-off these parameters from the pair $(𝒯,𝒜)$.
	
	The \emph{block length} $N$ of $𝒯$ is
	the number of instances of $W$ in the circuit.
	\[N≔2^{\depth(𝒯)}.\]
	
	The \emph{code rate} $R$ of $(𝒯,𝒜)$ is the number of instances of
	synthetic channels generated by the circuit that are included in $𝒜$,
	divided by the number of root channels the circuit consumes.
	It is also the probability of $K_τ$ ending up in $𝒜$.
	\[R≔ℙ\{K_τ∈𝒜\}.\]
	
	The \emph{error probability} $P$ of $(𝒯,𝒜)$ is
	the probability that any leaf channel in $𝒜$ fails to transmit the message.
	For the usual polar codes, this quantity is less than
	the weighted sum given in \cite[Proposition~2]{Arikan09}.
	\[P≤\sum_{w\in𝒜}Nℙ\{K_τ=w\}Z(w).\]
	This is still true in our case, proof omitted.
	
	The \emph{per-block time complexity}
	is the time the circuit generated by $𝒯$ takes to execute.
	It is bounded from above by the number of magic devices
	multiplied by the time each magic device spends.
	(No parallelism allowed).
	
	The reader can find in Appendix~\ref{app:auto} how magic devices work,
	verification omitted.
	The construction suggests that each magic device spends constant time.
	With the help of Appendix~\ref{app:polish}, the reader can also find
	that the total number of magic devices in the circuit is
	\[N𝔼[τ].\]
	(Hint: double-count the number of devices each wire passes.)
	Thus the per-block time complexity is proportional to
	\[N𝔼[τ].\]
	
	The \emph{per-bit time complexity} is
	the amortized time each information bit should pay.
	Unless the rate vanishes, it is proportional to
	\[𝔼[τ]. \label{eq:Etau}\]

\subsection{Grow a Tree}

	We have shown how to estimate the performance of a block code $(𝒯,𝒜)$
	if $𝒯$ and $𝒜$ are explicitly given.
	Now we demonstrate how to grow a good tree of prescribed depth $n$.
	
	Begin with $W$ as the only vertex of a new rooted tree.
	Let $Y(w)$ be $\min\{Z(w),1-Z(w)\}$ in the following framed rule:
	\TArule{
		Apply $T_\Ari$ to $w$ if and only if \\
		$\depth(w)<n$ and $Y(w)>ε2^{-n}$.
	}
	
	The rule says: for each leaf $w$,
	if the criteria $\depth(w)<n$ and $Y(w)>ε2^{-n}$ are met,
	apply $T_\Ari$ to $w$ to obtain $w^♭$ and $w^♯$
	(just like Formula~(\ref{eq:cir2}) and (\ref{eq:tre2}));
	and then append $w^♭$ and $w^♯$ as children of $w$.
	If, otherwise, either criterion is not met, leave $w$ as a leaf.
	
	This is a possible execution of the rule
	with $Z(W)=.5$,\ \ $n=3$,\ \ and $ε=.5$.
	Start with $W$ and write down $Z(W)$
	\[\tikz[ut]{
		\utnode$.5$${}{}
	}.\]
	Both $.5$ and $1-.5$ are larger than $.5\cdot2^{-3}$,
	so we append $1-(1-.5)^2$ and $.5^2$
	
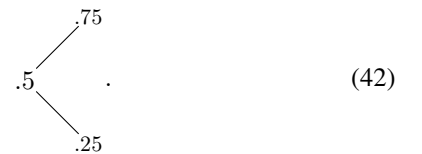
\[\tikz[ut]{
		\utnode$.5$${
			\utnode$.75$${}{} }{
			\utnode$.25$${}{} }
	}.\]
	Both $.75$ and $1-.75$ are larger than $.5\cdot2^{-3}$,
	so we append $1-(1-.75)^2$ and $.75^2$
	
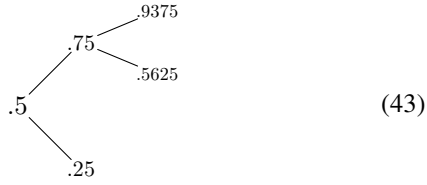
\[\tikz[ut]{
		\utnode$.5$${
			\utnode$.75$${
				\utnode$.9375$${}{} }{
				\utnode$.5625$${}{} } }{
			\utnode$.25$${}{} }
	}\]
	Both $.25$ and $1-.25$ are larger than $.5\cdot2^{-3}$,
	so we append $1-(1-.25)^2$ and $.25^2$
	
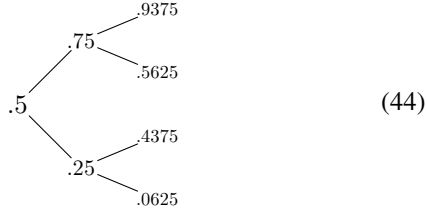
\[\tikz[ut]{
		\utnode$.5$${
			\utnode$.75$${
				\utnode$.9375$${}{} }{
				\utnode$.5625$${}{} } }{
			\utnode$.25$${
				\utnode$.4375$${}{} }{
				\utnode$.0625$${}{} } }
	}\]
	Among the four newcomers, the second and the third are such that
	$Y(w)>.5\cdot2^{-3}$, so we grow them further
	
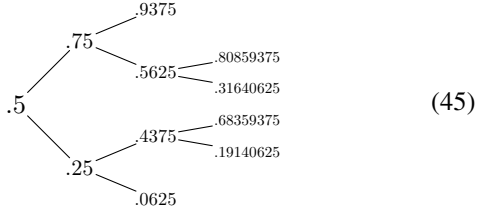
\[\tikz[ut]{
		\utnode$.5$${
			\utnode$.75$${
				\utnode$.9375$${}{} }{
				\utnode$.5625$${
					\utnode$.80859375$${}{} }{
					\utnode$.31640625$${}{} } } }{
			\utnode$.25$${
				\utnode$.4375$${
					\utnode$.68359375$${}{} }{
					\utnode$.19140625$${}{} } }{
				\utnode$.0625$${}{} } }
	}\]
	Now we reach depth $n=3$; terminate.
	See Appendix~\ref{app:prune} for another visualization.
	
	Having $𝒯$, we declare $𝒜$ by
	\TArule{
		$w∈𝒜$ if and only if \\
		$w$ is a leaf and $Z(w)≤ε2^{-n}$.
	}
	
	We show in the coming section how $(𝒯,𝒜)$ performs.

\section{Main Result} \label{sec:main}

	\def\mag#1{\hbox{\Large$\displaystyle#1$}}
	The following lemma is inspiring.
	\begin{lem}\cite[Theorem~1]{GX13}
		There exists $μ>0$ such that
		\[ℙ｛\mag{Z_i≤2^{-2^{.49i}}}｝≥I(W)-O(2^{-i/μ}).\]
	\end{lem}

	The following lemma generalizes the idea.
	\begin{lem}\cite[Theorem~3 and Formula~(56)]{MHU16}
		For $μ=3.627$ and $γ$ such that $1/(1+μ)<γ<1$,
		\begin{multline}
			ℙ｛\mag{Z_i≤2^{-2^{iγh_2^{-1}((γμ+γ-1)/γμ)}}}｝ \\
			≥I(W)-O(2^{-i(1-γ)/μ}).
		\end{multline}
	\end{lem}
	Here $h_2^{-1}$ is the inverse function of the binary entropy function.
	This lemma almost suffices for the choice of constants in this work.
	A stronger version of the lemma is in our previous work.
	\begin{lem}\label{lem:md}\cite[Theorem~6]{WD18}
		If for $π∈[0,1],$
		\[\frac{1-π}{μ'-μπ}+h_2（\frac{β'μ'}{μ'-μπ}）<1, \label{eq:picri}\]
		then
		\[ℙ｛\mag{Z_i≤2^{-2^{iβ'}}}｝≥I(W)-O(2^{-i/μ'}). \label{eq:betamu}\]
	\end{lem}
	
	For $μ=3.627$ given by \cite{FV14} and $(μ',β')=(4,1/24)$ chosen by us,
	Formula~(\ref{eq:picri}) becomes
	\[\frac4{4-3.627π}+h_2（\frac{1/6}{4-3.627π}）<1,\]
	which holds for all $π∈[0,1]$, as shown below.
	\pgfmathdeclarefunction{h2}{1}{\pgfmathparse{-#1*log2(#1)-(1-#1)*log2(1-#1)}}
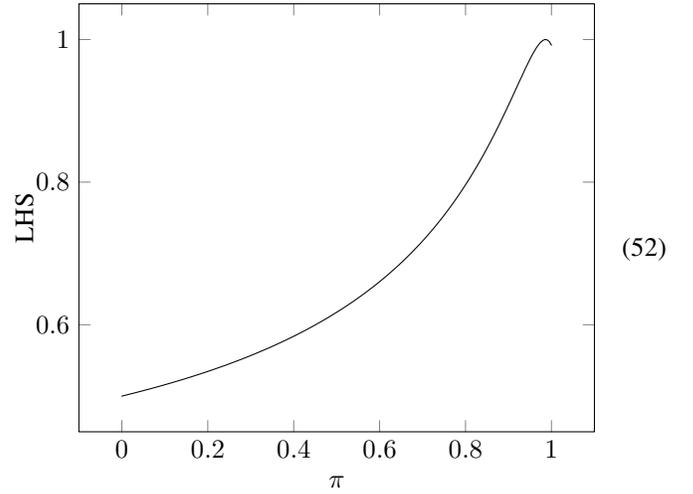
	\[\tikz{
		\begin{axis}[ylabel=LHS,
			         xlabel={$π$}]
			\addplot[domain=0:1,samples=200]
				{(1-\x)/(4-3.627*\x)+h2(1/(24-21.762*\x))};
		\end{axis}
	}\]
	Thus Formula~(\ref{eq:betamu}) becomes
	\[ℙ｛Z_i≤2^{-2^{i/24}}｝≥I(W)-O(2^{-i/4}). \label{eq:Z24I4}\]
	Since we are on erasure channels, the ``flipped version''
	\[ℙ｛I_i≤2^{-2^{i/24}}｝≥Z(W)-O(2^{-i/4}) \label{eq:I24Z4}\]
	also holds.
	
	We are now ready to state and prove the main theorem of this work.
	Recall that $Y(w)≔\min\{Z(w),1-Z(w)\}$.
	Let $Y_i$ be $\min\{Z_i,1-Z_i\}$.
	
	\begin{thm} \label{thm:main}
		Given $W$ and $ε$.
		Assign $n≔-5\log_2ε$.
		The framed rule
		\TArule{
			Apply $T_\Ari$ to $w$ if and only if \\
			$\depth(w)<n$ and $Y(w)>ε2^{-n}$.
		}
		generates a circuit with $O(N\loglogε)$ uses of magic devices.
	\end{thm}
	\begin{IEEEproof}
		Formulae (\ref{eq:Z24I4}) and~(\ref{eq:I24Z4})
		give the trichotomy and the corresponding probabilities:
		\[\begin{cases*}
			0≤Z_i≤2^{-2^{i/24}}                 & w.p.\ $I(W)-O(2^{-i/4})$; \\
			  2^{-2^{i/24}}<Z_i<1-2^{-2^{i/24}} & w.p.\ $     O(2^{-i/4})$; \\
			1-2^{-2^{i/24}}≤Z_i≤1               & w.p.\ $Z(W)-O(2^{-i/4})$.
		\end{cases*}\mskip-24mu\]
		Here ``w.p.'' reads ``with probability''.
		In terms of $Y_i$, the second line becomes
		\[ℙ｛Y_i>2^{-2^{i/24}}｝=O(2^{-i/4}). \label{eq:Y24O4}\]
		
		Let $τ$ be the stopping time
		\[τ≔\min（\{i\text{ such that }Y_i≤ε2^{-n}\}∪\{n\}）. \label{eq:stop}\]
		Then
		\begin{align}
			\{τ>i\}
			&⊂ \{Y_i>ε2^{-n}\} \\
			&= \{Y_i>ε^6\} \\
			&⊂ ｛Y_i>2^{-2^{i/24}}\text{ or }2^{-2^{i/24}}>ε^6｝ \\
			&⊂ ｛Y_i>2^{-2^{i/24}}\text{ or }i<O(\loglogε)｝.
		\end{align}
		The last line breaks into two cases:
		(a) when $i$ is large, i.e., when the second disjunct is false,
		the first disjunct must happen so Formula~(\ref{eq:Y24O4}) applies;
		(b) when $i$ is small, we expect no synthetic channel to be polarized
		so we apply the worst, yet educated bound, $1$.
		That is,
		\[ℙ\{τ>i\}≤\begin{cases*}
			O(2^{-i/4}) & when $i>O(\loglogε)$; \\
			1 & otherwise.
		\end{cases*}\]
		Therefore, we obtain an estimate by \cite[Lemma~2.2.8]{Durrett10}
		\[𝔼[τ]=∑_{i=0}^∞ℙ\{τ>i\}=O(\loglogε).\]
		
		Now generate a channel tree $𝒯$ with root $W$ by the framed rule.
		The criteria in the rule coincide with
		the stopping time $τ$ in Formula~(\ref{eq:stop}).
		Thus $\tau$ coincides with $\depth(K_τ)$,
		and the number of magic devices is bounded by
		\[N𝔼[τ]=O(N\loglogε). \label{eq:logloge}\]
	\end{IEEEproof}
	
	\begin{prop} \label{prop:N}
		The tree $𝒯$ defined above possesses block length $N=2^n=ε^{-5}$.
	\end{prop}
	\begin{IEEEproof}
		By the fact that the framed rule stops applying $T_\Ari$ at depth $n$.
	\end{IEEEproof}
	
	\begin{prop} \label{prop:C}
		The tree $𝒯$ defined above possesses per-bit time complexity
		$O(\loglogε)$.
	\end{prop}
	\begin{IEEEproof}
		By Formula~(\ref{eq:logloge}) and
		the discussion that leads to Formula~(\ref{eq:Etau}).
	\end{IEEEproof}
	
	\begin{prop} \label{prop:P}
		Given $𝒯$ defined above, define $𝒜$ by
		\TArule{
			$w∈𝒜$ if and only if \\
			$w$ is a leaf and $Z(w)≤ε2^{-n}$.
		}
		Then $(𝒯,𝒜)$ possesses block error probability $ε$.
	\end{prop}
	\begin{IEEEproof}
		Compute the error probability
		\begin{align}
			P
			&≤ \sum_{w\in𝒜}Nℙ\{K_τ=w\}Z(w) \\
			&≤ \sum_{w\in𝒜}Nℙ\{K_τ=w\}ε2^{-n} \\
			&= NRε2^{-n}\\
			&= Rε \\
			&≤ ε.
		\end{align}
	\end{IEEEproof}
	
	\begin{prop} \label{prop:R}
		The pair $(𝒯,𝒜)$ defined above possesses code rate $I(W)-ε$.
	\end{prop}
	\begin{IEEEproof}
		The sample space is partitioned into the following three events:
		\begin{align}
			S &≔ \{0≤Z_τ≤ε2^{-n}\}; \\
			M &≔ \{ε2^{-n}<Z_i<1-ε2^{-n}\text{ for all }i≤n\}; \\
			L &≔ \{1-ε2^{-n}≤Z_τ≤1\}.
		\end{align}
		
		Recall $n≔-5\log_2ε$.
		The second event is
		\begin{align}
			M
			&= \{Y_i>ε2^{-n}\text{ for all }i≤n\} \\
			&⊂ \{Y_n>ε2^{-n}\} \\
			&= \{Y_n>2^{-6/5n}\} \\
			&⊂ ｛Y_n>2^{-2^{n/24}}\text{ or }2^{-2^{n/24}}>2^{-6n/5}｝ \\
			&⊂ ｛Y_n>2^{-2^{n/24}}\text{ or }n<O(\log n)｝.
		\end{align}
		For $ε$ small enough ($n$ large enough),
		the case $n<O(\log n)$ does not happen.
		Thus whether $Y_n>2^{-2^{n/24}}$ happens dominants $M$.
		By Formula~(\ref{eq:Y24O4}),
		\[ℙ(M)≤O(2^{-n/4}).\]
		
		Rewrite the capacity; here $𝕀(\bullet)$ is the indicator function:
		\begin{align}
			I(W)
			&= 𝔼[I_τ] \\
			&= 𝔼[I_τ𝕀(S)]+𝔼[I_τ𝕀(M)]+𝔼[I_τ𝕀(L)] \\
			&= 𝔼[I_τ𝕀(S)]+𝔼[I_τ𝕀(M)]+𝔼[(1-Z_τ)𝕀(L)] \\
			&≤ 𝔼[𝕀(S)]+𝔼[𝕀(M)]+ε2^{-n}𝔼[𝕀(L)] \\
			&= ℙ(S)+ℙ(M)+ε2^{-n}ℙ(L) \\
			&≤ ℙ(S)+O(2^{-n/4})+ε2^{-n}.
		\end{align}
		Use it to compute the code rate:
		\begin{align}
			R
			&= ℙ\{K_τ∈𝒜\} \\
			&= ℙ(S) \\
			&≥ I(W)-O(2^{-n/4})-ε2^{-n} \\
			&= I(W)-O(ε^{5/4})-ε^6 \\
			&≥ I(W)-ε
		\end{align}
		for $ε$ small enough ($n$ large enough).
	\end{IEEEproof}
	
	Combining Proposition~\ref{prop:N},
	\ref{prop:C}, \ref{prop:P}, and \ref{prop:R},
	we certify that the constructed code $(𝒯,𝒜)$
	satisfies the properties claimed in the abstract.

\section{Connection to Other Works} \label{sec:connect}

\subsection{In Terms of Deleting Vertices}

	\cite{AYK11} introduces the so-called
	``simplified successive cancellation'' decoder, working as below:
	During the construction of polar codes,
	some synthetic channel, for instance $(W^♭)^♭$, may find that
	all its descendants are frozen (potentially because $(W^♭)^♭$ is too bad).
	In such case, it is unnecessary to establish
	the part of en-decoder circuit that corresponds to its children.
	
	Readers may find that the paragraph above coincides with
	the philosophy of Formula~(\ref{eq:tre7}) and (\ref{eq:cir7}).
	
	\cite{AYK11} then calls the synthetic channel $(W^♭)^♭$ a ``rate-zero node''.
	Similarly, a ``rate-one node'' is a synthetic channel that is so good,
	all of its descendants being utilized.
	In such case, \cite{AYK11} argues that it could save some time
	by shortcutting the classical successive cancelation decoder of \Arikan.
	
	That said, \cite{AYK11} does not realize that
	by not applying $T_\Ari$ in the first place it could have saved more,
	ultimately reducing the per-bit time complexity
	from $\log N$ to $\log\log N$.
	Frankly speaking, \cite{AYK11} is aiming for general channels
	while our result applies only to erasure channels.
	
	\cite{ZZWZP15} does similar things to polar codes with other kernels.
	\cite{ZZPYG14} does similar things, but is based on belief propagation.

\subsection{In Terms of Adding Vertices}

	\cite{EKMFLK15,EKMFLK17} introduce the so called ``relaxed polarization''.
	\cite{WLZZ15} introduces the so-called ``selective polarization''.
	They suggest that when some synthetic channel, say $(W^♭)^♯$,
	is not perfectly polarized,
	it should be further polarized by concatenating with an outer polar code.
	
	Readers may find that the paragraph above coincides with
	the philosophy of Formula~(\ref{eq:tre5}) and (\ref{eq:cir5}).
	
	\cite{EECB17} is another attempt, which they called ``code augmentation'',
	to protect unpolarized channels by appending polar codes to them.
	
	\cite{WYY18,WYXY18} do very similar things
	which they called ``information-coupling''.
	The idea is: some information bit might not be well-protected
	by the synthetic channel, say $(W^♭)^♯$, that it goes through.
	For the sake of reliability, send the same bit again
	through the same synthetic channel, $(W^♭)^♯$, in the very next block.
	Doing so merges two consecutive blocks into one big block.

\subsection{In Terms of Special Treatment}

	Recall the recursive definition
	\[Z_{i+1}=\begin{cases*}
		1-(1-Z_i)^2 & w.p.\ $1/2$ (head); \\
		Z_i^2       & w.p.\ $1/2$ (tail).
	\end{cases*}\]
	Assume for some $m∈[2n/5,n]$,
	\[ε2^{-n}<Z_m<ε2^{m-7n/5}.\]
	It is clear that
	although this synthetic channel is quite good,
	it is not good enough to become a leaf.
	What can we say about its descendants?
	
	Since $Z_{m+i+1}<2Z_{m+i}$, it turns out
	\[Z_{m+i}<2^iZ_m<ε2^{i+m-7n/5}<ε2^{-2n/5}\]
	for all $i<n-m$.
	Thus if tail ever happens, say at time $m+i+1$, then
	\[Z_{m+i+1}=Z_{m+i}^2<ε^22^{-4n/5}=ε2^{-n},\]
	which means a leaf.
	That is, the subtree rooted at $K_m$ is such that
	every down-child becomes a leaf,
	and every up-child has children, till depth $n$.
	Visually,
	
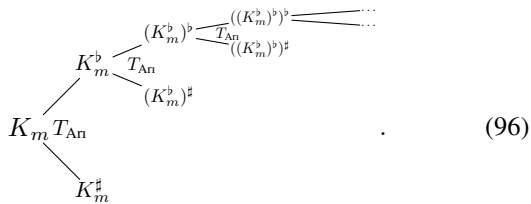
\[\tikz[ut]{
		\utnode$K_m$T_\Ari${
			\utnode$K_m^♭$T_\Ari${
				\utnode$(K_m^♭)^♭$T_\Ari${
					\utnode$((K_m^♭)^♭)^♭$${
						\utnode$\dotso$${}{} }{
						\utnode$\dotso$${}{} } }{
					\utnode$((K_m^♭)^♭)^♯$${}{} } }{
				\utnode$(K_m^♭)^♯$${}{} } }{
			\utnode$K_m^♯$${}{} }
	}.\]
	The upper-child at depth $n$ is then frozen
	while all other leaves are utilized.
	
	\cite{SG13} recognizes that this subtree
	generates a single-parity-check subcode,
	which can be decoded more efficiently than the magic devices do.
	
	Similarly, a $Z_m$ that is close enough to the top threshold $1-ε2^{-n}$
	generates a subtree that mainly ``grows downward''
	and every leaf except the very bottom one is frozen.
	Visually,
	
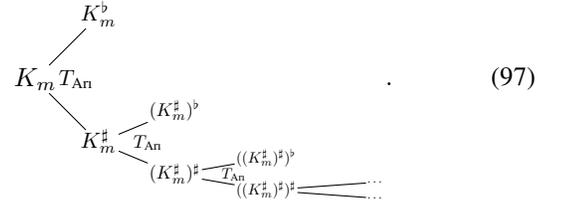
\[\tikz[ut]{
		\utnode$K_m$T_\Ari${
			\utnode$K_m^♭$${}{} }{
			\utnode$K_m^♯$T_\Ari${
				\utnode$(K_m^♯)^♭$${}{} }{
				\utnode$(K_m^♯)^♯$T_\Ari${
					\utnode$((K_m^♯)^♯)^♭$${}{} }{
					\utnode$((K_m^♯)^♯)^♯$${
						\utnode$\dotso$${}{} }{
						\utnode$\dotso$${}{} } } } }
	}.\]
	This either induces a trivial code (if the very bottom leaf is frozen)
	or a repetition code (if the very bottom leaf is utilized)
	and, again, can be efficiently decoded.
	
	The simulation by \cite{SG13}, and subsequently by \cite{SGVTG14},
	suggests that this \latin{ad hoc} treatment
	accelerates the real world performance.
	For our purpose, however,
	special treatment makes it difficult
	to bound the complexity as they are special.

\subsection{In Terms of Systematic Coding}

	\cite{Arikan11} suggests systematic polar coding,
	where the receiver is not interested in $ˆu$
	but wants to recover $x$ from $y$.
	
	One consequence is that, if the two right pins of the magic device
	\[\tikz\pic{D};\]
	correspond to two frozen channels,
	then this device can be dropped
	without affecting the overall decoding ability of the circuit.
	Similarly, if the two right pins correspond to two utilized channels,
	it could also be dropped.
	
	The argument above gives another reason (or perspective)
	why the tree should be pruned.
	One may keep dropping magic devices (keep pruning the tree)
	till it stabilizes.
	It is easy to see that a device remains if and only if
	some of its children are frozen and some are utilized.
	Then it is not hard to estimate the number of remaining devices.
	
	Our intuition suggests that the number of remaining devices is
	\[O（N\log\left\lvert\log\frac{ϵ}N\right\rvert）\]
	where $ϵ$ is the threshold of a channel being utilized
	(which is $ε2^{-n}$ in our construction).
	When $N$ is polynomial in $ϵ$, this reassures out result.
	This is the strategy we refer to in the second paragraph of the abstract.

\section{Future Works}

	For more general binary channels such as BSC or BI-AWGN,
	Formula~(\ref{eq:I24Z4}) is no longer true.
	Consequently a large portion of estimation done in this work does not apply.
	Potentially one can mimic \cite[Appendix~A]{MHU16}
	to control $I(W)-R$.
	
	From studies of random codes,
	$I(W)-R$ is polynomial in $N$ while $P$ is exponential in $N$.
	Thus it seems improper
	to parametrize $I(W)-R$ and $P$ with a single variable $ε$.
	It would be interesting if one could come up with a description
	of more general tradeoffs among $N$, $R$, $P$, and time complexity.

\section{Concluding Remarks}

	We propose a pruned variant of polar coding
	where the channel tree is pruned by closely looking at the \Bha{} parameters.
	Then we prove that the per-bit complexity is log-logarithmic
	in block length, in gap to capacity and in error probability.
	
	This idea turns out
	to coincide with some existing works mentioned in Section~\ref{sec:connect}.
	They found that doing this type of simplification
	reduces the empirical execution time significantly.
	But we could not find any statement about the log-logarithmic asymptote.
	(\Arikan{} mentions time complexity $O(N\log N)$
	in \cite[Section~VII-C, last paragraph]{Arikan16}.)
	
	Although the log-logarithmic asymptote is not record-breaking as
	other constructions with bounded per-bit complexity exist \cite{PSU04,PS05},
	our construction controls the block length while other works do not.

{
	\let\bobitem\bibitem
	\def\bibitem[#1]#2{
		\def\newpageat{EKMFLK15}
		\def\thisitem{#2}
		\ifx
			\newpageat\thisitem\newpage
			\linespread{1.036}\selectfont
		\fi
		\bobitem[#1]{#2}
	}
	\bibliographystyle{alpha}
	\bibliography{LoglogTime-4}
}

\clearpage

\appendix

\subsection{Simulation}

	We wrote a python script to support our claims.
	The script chooses $I(W)=.618$, varies $n$, calculates $ε=2^{-n/5}$,
	and profiles the process of \SI{1}{\mebi\bit}.
	The per-bit time is shown below.
	
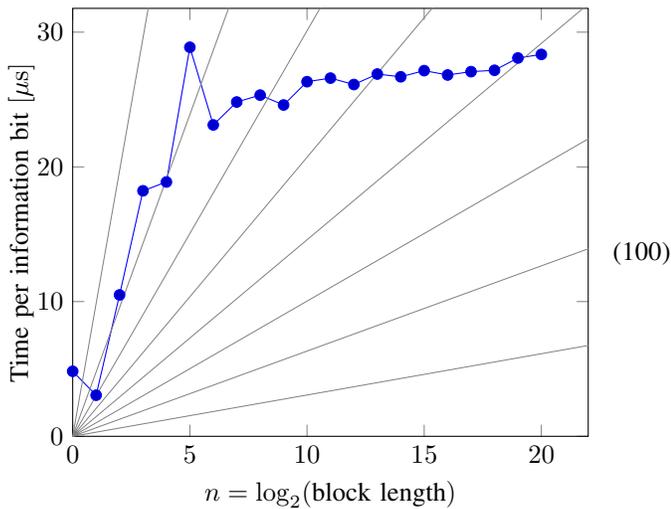
\[\tikz{
		\begin{axis}[simu,
			         ylabel=Time per information bit,
			         change y base,y SI prefix=micro,y unit=s,
			         xlabel={$n=\log_2(\text{block length})$}]
			\addplot table{
				0	4.82528412249e-06
				1	3.04982745547e-06
				2	1.04944288069e-05
				3	1.8223561104e-05
				4	1.88777563987e-05
				5	2.88788760372e-05
				6	2.31143033792e-05
				7	2.48139676842e-05
				8	2.53270508256e-05
				9	2.45934966693e-05
				10	2.63224969136e-05
				11	2.65883469673e-05
				12	2.61120344541e-05
				13	2.68892748458e-05
				14	2.66861428305e-05
				15	2.7144143333e-05
				16	2.68208824836e-05
				17	2.7068153533e-05
				18	2.71689202457e-05
				19	2.80819817421e-05
				20	2.83468062578e-05
			};
			\draw[hl]foreach\i in{10,20,...,80}{(0,0)--(\i:10cm)};
		\end{axis}
	}\]
	Notice how the per-bit time \emph{does not} grow proportionally to $n$,
	while it does in classical polar codes.
	
	The plot does not really prove anything as there are a lot of factors.
	For instance:
	(a) it is python;
	(b) it tests a small amount of data;
	(c) the channel is simulated by the built-in PRNG,
		which might dominate the performance;
	(d) the tree traversal is implemented by function recursion,
		wherein a function call serves as few as one bit
		for leaves at the very bottom of the tree.
	
	It is more obvious if we look directly at $𝔼[τ]$
	(and believe that
	the real world performance is really proportional to $𝔼[τ]$).
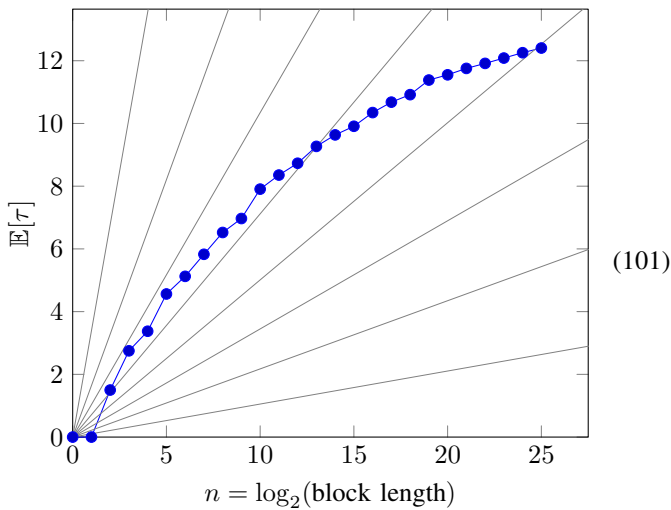
	\[\tikz{
		\begin{axis}[simu,
			         ylabel={$𝔼[τ]$},
			         xlabel={$n=\log_2(\text{block length})$}]
			\addplot table{
				0	0.0
				1	0.0
				2	1.5
				3	2.75
				4	3.375
				5	4.5625
				6	5.125
				7	5.828125
				8	6.5234375
				9	6.96875
				10	7.90625
				11	8.3544921875
				12	8.7314453125
				13	9.26879882812
				14	9.63500976562
				15	9.91046142578
				16	10.3467407227
				17	10.6794891357
				18	10.9180297852
				19	11.3841133118
				20	11.5504550934
				21	11.7551832199
				22	11.9149127007
				23	12.0850632191
				24	12.255657196
				25	12.4046368599
			};
			\draw[hl]foreach\i in{10,20,...,80}{(0,0)--(\i:10cm)};
		\end{axis}
	}\]
	Notice how the plot bends downward like $\loglogε≈\log n$ does.
	At $n=25$, our construction saves half of magic devices.
	
	Starting from $n=26$ it is difficult to calculate the exact value of $𝔼[τ]$.
	We instead sample the process $Z_n$ a thousand times.
	The result is majestic.
	
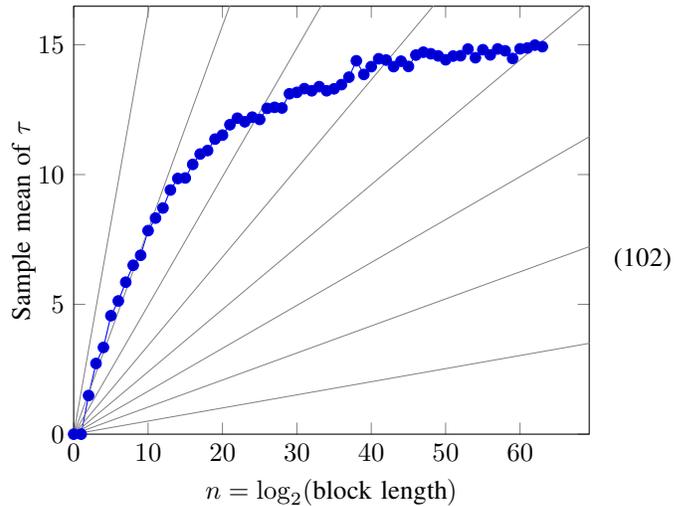
\[\tikz{
		\begin{axis}[simu,
			         ylabel={Sample mean of $\tau$},
			         xlabel={$n=\log_2(\text{block length})$}]
			\addplot table{
				0	0.0
				1	0.0
				2	1.489
				3	2.723
				4	3.341
				5	4.563
				6	5.132
				7	5.854
				8	6.506
				9	6.89
				10	7.84
				11	8.324
				12	8.712
				13	9.405
				14	9.849
				15	9.869
				16	10.391
				17	10.791
				18	10.926
				19	11.366
				20	11.52
				21	11.92
				22	12.171
				23	12.033
				24	12.211
				25	12.123
				26	12.55
				27	12.59
				28	12.564
				29	13.112
				30	13.166
				31	13.31
				32	13.227
				33	13.387
				34	13.228
				35	13.303
				36	13.465
				37	13.751
				38	14.383
				39	13.857
				40	14.158
				41	14.461
				42	14.412
				43	14.158
				44	14.367
				45	14.165
				46	14.603
				47	14.714
				48	14.654
				49	14.572
				50	14.422
				51	14.565
				52	14.578
				53	14.834
				54	14.501
				55	14.81
				56	14.611
				57	14.84
				58	14.767
				59	14.472
				60	14.844
				61	14.876
				62	14.986
				63	14.927
			};
			\draw[hl]foreach\i in{10,20,...,80}{(0,0)--(\i:10cm)};
		\end{axis}
	}\]
	The plot stops at $n=63$ because that is about the size of the internet,
	where the sample mean of $τ$ does not exceed $16$.
	The sample mean of $τ$ exceeds $17$ when $n≈240$.
	That is, when the block length
	is about the number of atoms in the known universe,
	and the error probability is $2^{-48}$.

\subsection{Gallery of Trees and Circuits} \label{app:polish}

	For trees, the labels of transformations and channels are omitted.
	
	For circuits, only the decoder component is shown;
	the encoder component is the reflection of the decoder component.
	Plus, we do not wire shallow channels to the right boundary as
	the order of shallow channels at a deeper layer is irrelevant.

	\makeatletter
	
	\def\utdot#1#2{
		\edef\utheight{\the\numexpr\utheight-1}
		\utmiddle.5\utupper\advance\utmiddle.5\utlower
		\fill(\utdepth,\utmiddle)circle(2^\utheight*.8pt)coordinate(\utdepth);
		\ifnum\utdepth>0
			\draw[line width=2^\utheight*.8](\the\numexpr\utdepth-1)--(\utdepth);
		\fi
		\edef\utdepth{\the\numexpr\utdepth+1}
		{\utlower\utmiddle#1}{\utupper\utmiddle#2}
	}
	
	\def\bbsetupcs{
		\def\pgfpointxy##1##2{
			\pgfpoint
				{2\bbwire*##1+xoffset(\bbxmax-1)-xoffset(\bbxmax-##1)}
				{(##2)*2\bbwire}
		}
	}
	\def\bbdepth{0}
	\def\bbprefix{}
	\def\bbinfix{0}
	\def\bbnode#1#2{
		\edef\bbheight{\the\numexpr\bbheight-1}
		\pgfmathtruncatemacro\bbtwotoh{2^\bbheight}
		\foreach\i in{1,...,\bbtwotoh}{
			\pgfmathsetmacro\bbpostfix{bin(\bbtwotoh+\i-1)}
			\edef\bbpostfix{\expandafter\pgfutil@gobble\bbpostfix}
			\draw(\bbdepth,0b\bbprefix\bbinfix*\bbtwotoh+\i)
			             pic(\bbprefix\bbinfix-\bbpostfix){D};
			\ifx\bbprefix\empty
			\else\draw
			(\bbprefix-\bbpostfix1\bbinfix R)--(\bbprefix\bbinfix-\bbpostfix1L)
			(\bbprefix-\bbpostfix0\bbinfix R)--(\bbprefix\bbinfix-\bbpostfix0L);
			\fi
		}
		\edef\bbprefix{\bbprefix\bbinfix}
		\edef\bbdepth{\the\numexpr\bbdepth+1}
		{\def\bbinfix{1}#1}{\def\bbinfix{0}#2}
	}
	\def\bbend#1#2{
		\edef\bbheight{\the\numexpr\bbheight-1}
		\pgfmathtruncatemacro\bbtwotoh{2^\bbheight}
		\ifnum\bbheight<0\else
		\foreach\i in{1,...,\bbtwotoh}{
			\pgfmathsetmacro\bbpostfix{bin(\bbtwotoh+\i-1)}
			\edef\bbpostfix{\expandafter\pgfutil@gobble\bbpostfix}
			\draw(\bbdepth,0b\bbprefix\bbinfix*\bbtwotoh+\i)
			             pic(\bbprefix\bbinfix-\bbpostfix){G};
			\draw
			(\bbprefix-\bbpostfix1\bbinfix R)--(\bbprefix\bbinfix-\bbpostfix1L)
			(\bbprefix-\bbpostfix0\bbinfix R)--(\bbprefix\bbinfix-\bbpostfix0L);
		}
		\fi
	}
	
	\def\popush#1{
		\let\EA\expandafter\let\postuck\postack
		\EA\gdef\EA\postack\EA{\EA#1\postuck}
	}
	\def\popopone#1#2\postackbuttom{\gdef\postack{#2}#1}
	\def\popop{\expandafter\popopone\postack\postackbuttom}
	\def\poreadone#1{%
		\if#10
			\g@addto@macro\poqueue{\poleaf{}{}}
		\else
			\popush\poscape
			\popush\pochildjunior
			\popush\pochildsenior
		\fi
		\popop
	}
	\def\pochildsenior{\g@addto@macro\poqueue{\pobin\{}\poreadone}
	\def\pochildjunior{\g@addto@macro\poqueue{\}\{}\poreadone}
	\def\poscape{\g@addto@macro\poqueue{\}}\popop}
	
	\def\polish#1{
		\def\postack{\relax}\def\poqueue{}
		\poreadone #1
		\def\{{\if22{\else}\fi}\def\}{\if23{\else}\fi}\let\NE\noexpand
		\def\pobin{\NE\utdot }\def\poleaf{\NE\utdot}\xdef\utqueue{\poqueue}
		\def\pobin{\NE\bbnode}\def\poleaf{\NE\bbend}\xdef\bbqueue{\poqueue}
		\let\bbheight\bbxmax
		\let\utheight\bbxmax
		\pgfmathsetlength\utupper{2^\bbxmax*\bbwire}
		\[\llap{\tikz[ut]{\utqueue}}=\rlap{\tikz[bb]{\bbqueue}}\]
		\message{polish}
	}

\bgroup
\advance\parskip\fill

\if00

	\def\bbxmax{0}
	\polish{
		0
	}
	
	\def\bbxmax{1}
	\polish{
		2
			0
			0
	}
	
	\def\bbxmax{2}
	\polish{
		2
			2
				0
				0
			0
	}
	
	\polish{
		2
			2
				0
				0
			2
				0
				0
	}
	
	\polish{
		2
			0
			2
				0
				0
	}
	
	\def\bbxmax{3}
	\polish{
		2
			2
				0
				2
					0
					0
			0
	}
	
	\polish{
		2
			2
				2
					0
					0
				2
					0
					0
			0
	}
	
	\polish{
		2
			2
				2
					0
					0
				0
			0
	}
	
	\polish{
		2
			2
				2
					0
					0
				0
			2
				0
				0
	}
	
	\polish{
		2
			2
				2
					0
					0
				2
					0
					0
			2
				0
				0
	}
	
	\polish{
		2
			2
				0
				2
					0
					0
			2
				0
				0
	}

\newpage

	\polish{
		2
			2
				0
				2
					0
					0
			2
				2
					0
					0
				0
	}
	
	\polish{
		2
			2
				2
					0
					0
				2
					0
					0
			2
				2
					0
					0
				0
	}
	
	\polish{
		2
			2
				2
					0
					0
				0
			2
				2
					0
					0
				0
	}
	
	\polish{
		2
			2
				0
				0
			2
				2
					0
					0
				0
	}
	
	\polish{
		2
			2
				0
				0
			2
				2
					0
					0
				2
					0
					0
	}

\newpage

	\polish{
		2
			2
				2
					0
					0
				0
			2
				2
					0
					0
				2
					0
					0
	}
	
	\polish{
		2
			2
				2
					0
					0
				2
					0
					0
			2
				2
					0
					0
				2
					0
					0
	}
	
	\polish{
		2
			2
				0
				2
					0
					0
			2
				2
					0
					0
				2
					0
					0
	}
	
	\polish{
		2
			2
				0
				2
					0
					0
			2
				0
				2
					0
					0
	}
	
	\polish{
		2
			2
				2
					0
					0
				2
					0
					0
			2
				0
				2
					0
					0
	}

\newpage

	\polish{
		2
			2
				2
					0
					0
				0
			2
				0
				2
					0
					0
	}
	
	\polish{
		2
			2
				0
				0
			2
				0
				2
					0
					0
	}
	
	\polish{
		2
			0
			2
				0
				2
					0
					0
	}
	
	\polish{
		2
			0
			2
				2
					0
					0
				2
					0
					0
	}
	
	\polish{
		2
			0
			2
				2
					0
					0
				0
	}

\onecolumn

	\def\bbxmax{4}
	\polish{
		2
			2
				0
				2
					2
						0
						0
					0
			0
	}
	
	\polish{
		2
			2
				0
				2
					0
					2
						0
						0
			0
	}
	
	\polish{
		2
			2
				2
					0
					2
						0
						0
				0
			0
	}

\newpage

	\polish{
		2
			0
			2
				2
					0
					2
						0
						0
				0
	}
	
	\def\bbxmax{5}
	\polish{
		2
			2
				0
				2
					2
						0
						2
							0
							0
					0
			0
	}

\fi

\egroup

\twocolumn

\linespread{1.009}\selectfont

\subsection{The Automata Model} \label{app:auto}

	\tikzset{
		AM/.style={nodes={text height=6,text depth=1}},
		FF/.pic={\def\coor{coordinate}\draw[pic actions]
			(-2 \bbwire, 2 \bbgape)\coor(1L)--(-2 \bbsize, 2 \bbgape)\coor(1l)--
			( 2 \bbsize, 2 \bbgape)\coor(1r)--( 2 \bbwire, 2 \bbgape)\coor(1R)
			(-2 \bbwire,-2 \bbgape)\coor(0L)--(-2 \bbsize,-2 \bbgape)\coor(0l)--
			( 2 \bbsize,-2 \bbgape)\coor(0r)--( 2 \bbwire,-2 \bbgape)\coor(0R);},
		EE/.pic={\pic[thick]{FF};\draw[thick,fill=encolor]
			(-2 \bbsize,-2 \bbsize)rectangle  ( 2 \bbsize, 2 \bbsize)
			(-  \bbgape,-  \bbgape)         --(   \bbgape,   \bbgape);},
		DD/.pic={\pic[thick]{FF};\draw[thick,fill=decolor]
			(-2 \bbsize,-2 \bbsize)rectangle  ( 2 \bbsize, 2 \bbsize)
			(-  \bbgape,   \bbgape)         --(   \bbgape,-  \bbgape)..controls
			( .5\bbgape,-.5\bbgape)    and    ( .5\bbgape, .5\bbgape)..
			(   \bbgape,   \bbgape)         --(-  \bbgape,-  \bbgape);},
		AC/.pic={\def\coord{coordinate}\draw[thick]
			(   \bbsize, 0        )         --( 2 \bbwire, 0        )\coord(R);},
		CB/.pic={\def\coord{coordinate}\draw[thick]
			(-2 \bbwire, 0        )\coord(L)--(-  \bbsize, 0        )         ;},
		AB/.pic={\draw[thick]
			(-  \bbsize,-  \bbsize)rectangle  (   \bbsize,   \bbsize);},
		CC/.pic={\pic{AC}pic{AB}pic{CB}node{$W$};},
		BB/.pic={\pic       {AB}pic{CB}         ;},
		AA/.pic={\pic{AC}pic{AB}                ;},
	}
	\def\bbstate#1#2#3#4#5#6#7#8#9{
		\tikz[AM]{
			\pic(){#1};
			\path[o]
		(1L)node[<<]{$#2$}(1l)node[>>]{$#3$}(1r)node[<<]{$#4$}(1R)node[>>]{$#5$}
		(0L)node[<<]{$#6$}(0l)node[>>]{$#7$}(0r)node[<<]{$#8$}(0R)node[>>]{$#9$};
		}
	}
	\def\bbestate{\bbstate{EE}}
	\def\bbdstate{\bbstate{DD}}
	\def\alistate#1#2{
		\tikz[AM]\pic(){AA}(0,0)node{$#1$}(R)node[>>]{$#2$};
	}
	\def\bobstate#1#2{
		\tikz[AM]\pic(){BB}(L)node[<<]{$#1$}(0,0)node{$#2$};
	}
	\def\chastate#1#2{
		\tikz[AM]\pic(){CC}(L)node[<<]{$#1$}(R)node[>>]{$#2$};
	}
	\def\State#1\becomes#2;{
		\[#1\quad\text{becomes}\quad#2\]
	}

\subsubsection{The Sending Component}

	The person who sends owns two types of devices.
	The frozen bit sender corresponds to frozen bits in polar coding context.
	It sends out zero, once.
	The utilized bit sender, on the other hand,
	sends the information bit the person wants to send, once.
	It is important that
	these automata send out bits once and become ``idle'' thereafter,
	so the circuit is not flooded by repetitive bits.
	\State  \alistate{\text F}{}
	\becomes\alistate{\text f}{0→};
	\State  \alistate{u}{}
	\becomes\alistate{}{u→};
	\State  \alistate{u}{}
	\becomes\alistate{}{u→};

\subsubsection{The Encoding Component}

	The encoding component consists of many many $(u+v,v)$-construction devices.
	The arithmetics is done in $𝔽_2$, i.e., $\operatorname{GF}(2)$.
	Every encoding device will be executed exactly once.
	\State  \bbestate{u→}{}{}{}
	                 {v→}{}{}{}
	\becomes\bbestate{}{}{}{u+v→}
	                 {}{}{}{v→}\quad;

\subsubsection{The Channel Component}

	We consider binary erasure channels.
	They are best described by the so called probabilistic automata.
	\def\?{\texttt?}
	\State  \chastate{x→}{}
	\becomes\chastate{}{x→};
	with probability $I(W)$; or
	\State  \phantom{\chastate{x→}{}}
	\becomes\chastate{}{\?→};
	with probability $Z(W)$.

\linespread{1}\selectfont

\vskip\belowdisplayskip

\subsubsection{The Decoding Component}

	The decoding component consists of devices that
	reverse the $(u+v,v)$-construction.
	Each decoding device is executed for three things:
	\vadjust{\newpage}
	Firstly it is activated by two incoming bits.
	It then tells the upper successor its best guess $ˆu=y-z$,
	which is supposed to be the $u$ given that inputs are $(u+v,v)$.
	Secondly it will receive feedback from the upper successor,
	saying that the correct bit is indeed $ˆu$.
	Based on this information,
	it tells the lower successor its best guess of $v$.
	Thirdly after the lower successor confirms the value of $ˆv$,
	it forwards the information it collects,
	in the form of $(u+v,v)$, to its predecessors.
	Here the subtraction involving $\?$ results in $\?$.
	The binary operator $∨$
	returns one of its non-$\?$ operand(s), if any; otherwise it returns $\?$.
	\State  \bbdstate{y→}{}{}{}
	                 {z→}{}{}{}
	\becomes\bbdstate{}{y}{}{y-z→}
	                 {}{z}{}{}\qquad;
	\State  \bbdstate{}{y}{}{←ˆu}
	                 {}{z}{}{}
	\becomes\bbdstate{}{}{ˆu}{}
	                 {}{}{}{(y-ˆu)∨z→}\qquad;
	\State  \bbdstate{}{}{ˆu}{}
	                 {}{}{}{←ˆv}
	\becomes\bbdstate{←ˆu+ˆv}{}{}{}
	                 {←ˆv}{}{}{}\qquad;

\subsubsection{The Receiving Component}

	The person who receives owns two type of devices.
	The frozen bit receiver receives any input symbol and reply $0$.
	The utilized bit receiver receives any input symbol
	and blindly replies the exact same symbol.
	If it happens that the utilized bit receiver receives $\?$,
	then there is no chance to recover this erasure anymore;
	the receiver may throw a \texttt{BlockError} exception
	that terminates the automata.
	\State  \bobstate{ˆu→}{\text F}
	\becomes\bobstate{←0}{\text f};
	\State  \bobstate{ˆu→}{}
	\becomes\bobstate{←ˆu}{ˆu};

\subsubsection{Assembled Automata}

	In the next page, notice that Formulae (\ref{eq:200}) and~(\ref{eq:2200200})
	are different codes, but provide the same protection.

\clearpage

\onecolumn

	\[\tikz[AM]{\draw
		(-10\bbwire,3  \bbsize)pic(A1){AA}node{F}
		(-10\bbwire,0  \bbsize)pic(A0){AA}node{F}
		(-5 \bbwire,1.5\bbsize)pic(E0){EE}
		( 0        ,3  \bbsize)pic(C1){CC}
		( 0        ,0  \bbsize)pic(C0){CC}
		( 5 \bbwire,1.5\bbsize)pic(D0){DD}
		( 10\bbwire,3  \bbsize)pic(B1){BB}node{F}
		( 10\bbwire,0  \bbsize)pic(B0){BB}node{F}
		(A1R)--(E01L)(E01R)--(C1L)(C1R)--(D01L)(D01R)--(B1L)
		(A0R)--(E00L)(E00R)--(C0L)(C0R)--(D00L)(D00R)--(B0L)
	}\]\vfil
	
	\[\tikz[AM]{\draw
		(-10\bbwire,3  \bbsize)pic(A1){AA}node{F}
		(-10\bbwire,0  \bbsize)pic(A0){AA}node{$u_1$}
		(-5 \bbwire,1.5\bbsize)pic(E0){EE}
		( 0        ,3  \bbsize)pic(C1){CC}
		( 0        ,0  \bbsize)pic(C0){CC}
		( 5 \bbwire,1.5\bbsize)pic(D0){DD}
		( 10\bbwire,3  \bbsize)pic(B1){BB}node{F}
		( 10\bbwire,0  \bbsize)pic(B0){BB}
		(A1R)--(E01L)(E01R)--(C1L)(C1R)--(D01L)(D01R)--(B1L)
		(A0R)--(E00L)(E00R)--(C0L)(C0R)--(D00L)(D00R)--(B0L)
	}\label{eq:200}\]\vfil
	
	\[\tikz[AM]{\draw
		(-10\bbwire,3  \bbsize)pic(A1){AA}node{$u_0$}
		(-10\bbwire,0  \bbsize)pic(A0){AA}node{$u_1$}
		(-5 \bbwire,1.5\bbsize)pic(E0){EE}
		( 0        ,3  \bbsize)pic(C1){CC}
		( 0        ,0  \bbsize)pic(C0){CC}
		( 5 \bbwire,1.5\bbsize)pic(D0){DD}
		( 10\bbwire,3  \bbsize)pic(B1){BB}
		( 10\bbwire,0  \bbsize)pic(B0){BB}
		(A1R)--(E01L)(E01R)--(C1L)
		                                 (C1R)--(D01L)(D01R)--(B1L)
		(A0R)--(E00L)(E00R)--(C0L)
		                                 (C0R)--(D00L)(D00R)--(B0L)
	}\]\vfil
	
	\[\tikz[AM]{\draw
		(-15\bbwire,9  \bbsize)pic(A3){AA}node{F}
		(-15\bbwire,6  \bbsize)pic(A2){AA}node{F}
		(-15\bbwire,3  \bbsize)pic(A1){AA}node{F}
		(-15\bbwire,0  \bbsize)pic(A0){AA}node{$u_3$}
		(-10\bbwire,7.5\bbsize)pic(E3){FF}(-5\bbwire,7.5\bbsize)pic(E1){EE}
		(-10\bbwire,1.5\bbsize)pic(E2){EE}(-5\bbwire,1.5\bbsize)pic(E0){EE}
		foreach\i in{0,1,2,3}{(0,3*\i\bbsize)pic(C\i){CC}}
		(  5\bbwire,7.5\bbsize)pic(D1){DD}(10\bbwire,7.5\bbsize)pic(D3){FF}
		(  5\bbwire,1.5\bbsize)pic(D0){DD}(10\bbwire,1.5\bbsize)pic(D2){DD}
		( 15\bbwire,9  \bbsize)pic(B3){BB}node{F}
		( 15\bbwire,6  \bbsize)pic(B2){BB}node{F}
		( 15\bbwire,3  \bbsize)pic(B1){BB}node{F}
		( 15\bbwire,0  \bbsize)pic(B0){BB}
		(A3R)--(E31L)(E31R)--(E11L)(E11R)--(C3L)
		                            (C3R)--(D11L)(D11R)--(D31L)(D31R)--(B3L)
		(A1R)--(E21L)(E21R)--(E10L)(E10R)--(C2L)
		                            (C2R)--(D10L)(D10R)--(D21L)(D21R)--(B1L)
		(A2R)--(E30L)(E30R)--(E01L)(E01R)--(C1L)
		                            (C1R)--(D01L)(D01R)--(D30L)(D30R)--(B2L)
		(A0R)--(E20L)(E20R)--(E00L)(E00R)--(C0L)
		                            (C0R)--(D00L)(D00R)--(D20L)(D20R)--(B0L)
	}\]\vfil
	
	\[\tikz[AM]{\draw
		(-15\bbwire,9  \bbsize)pic(A3){AA}node{F}
		(-15\bbwire,6  \bbsize)pic(A2){AA}node{F}
		(-15\bbwire,3  \bbsize)pic(A1){AA}node{$u_2$}
		(-15\bbwire,0  \bbsize)pic(A0){AA}node{$u_3$}
		(-10\bbwire,7.5\bbsize)pic(E3){EE}(-5\bbwire,7.5\bbsize)pic(E1){EE}
		(-10\bbwire,1.5\bbsize)pic(E2){EE}(-5\bbwire,1.5\bbsize)pic(E0){EE}
		foreach\i in{0,1,2,3}{(0,3*\i\bbsize)pic(C\i){CC}}
		(  5\bbwire,7.5\bbsize)pic(D1){DD}(10\bbwire,7.5\bbsize)pic(D3){DD}
		(  5\bbwire,1.5\bbsize)pic(D0){DD}(10\bbwire,1.5\bbsize)pic(D2){DD}
		( 15\bbwire,9  \bbsize)pic(B3){BB}node{F}
		( 15\bbwire,6  \bbsize)pic(B2){BB}node{F}
		( 15\bbwire,3  \bbsize)pic(B1){BB}
		( 15\bbwire,0  \bbsize)pic(B0){BB}
		(A3R)--(E31L)(E31R)--(E11L)(E11R)--(C3L)
		                            (C3R)--(D11L)(D11R)--(D31L)(D31R)--(B3L)
		(A1R)--(E21L)(E21R)--(E10L)(E10R)--(C2L)
		                            (C2R)--(D10L)(D10R)--(D21L)(D21R)--(B1L)
		(A2R)--(E30L)(E30R)--(E01L)(E01R)--(C1L)
		                            (C1R)--(D01L)(D01R)--(D30L)(D30R)--(B2L)
		(A0R)--(E20L)(E20R)--(E00L)(E00R)--(C0L)
		                            (C0R)--(D00L)(D00R)--(D20L)(D20R)--(B0L)
	}\label{eq:2200200}\]\vfil
	
	\[\tikz[AM]{\draw
		(-15\bbwire,9  \bbsize)pic(A3){AA}node{F}
		(-15\bbwire,6  \bbsize)pic(A2){AA}node{$u_1$}
		(-15\bbwire,3  \bbsize)pic(A1){AA}node{$u_2$}
		(-15\bbwire,0  \bbsize)pic(A0){AA}node{$u_3$}
		(-10\bbwire,7.5\bbsize)pic(E3){EE}(-5\bbwire,7.5\bbsize)pic(E1){EE}
		(-10\bbwire,1.5\bbsize)pic(E2){FF}(-5\bbwire,1.5\bbsize)pic(E0){EE}
		foreach\i in{0,1,2,3}{(0,3*\i\bbsize)pic(C\i){CC}}
		(  5\bbwire,7.5\bbsize)pic(D1){DD}(10\bbwire,7.5\bbsize)pic(D3){DD}
		(  5\bbwire,1.5\bbsize)pic(D0){DD}(10\bbwire,1.5\bbsize)pic(D2){FF}
		( 15\bbwire,9  \bbsize)pic(B3){BB}node{F}
		( 15\bbwire,6  \bbsize)pic(B2){BB}
		( 15\bbwire,3  \bbsize)pic(B1){BB}
		( 15\bbwire,0  \bbsize)pic(B0){BB}
		(A3R)--(E31L)(E31R)--(E11L)(E11R)--(C3L)
		                            (C3R)--(D11L)(D11R)--(D31L)(D31R)--(B3L)
		(A1R)--(E21L)(E21R)--(E10L)(E10R)--(C2L)
		                            (C2R)--(D10L)(D10R)--(D21L)(D21R)--(B1L)
		(A2R)--(E30L)(E30R)--(E01L)(E01R)--(C1L)
		                            (C1R)--(D01L)(D01R)--(D30L)(D30R)--(B2L)
		(A0R)--(E20L)(E20R)--(E00L)(E00R)--(C0L)
		                            (C0R)--(D00L)(D00R)--(D20L)(D20R)--(B0L)
	}\]\vfil

\vfilneg\vfilneg
\clearpage

\subsection{Growing/Pruning Visualization} \label{app:prune}

	For each segment, the line width is $\sqrt2$ thinner than its predecessor,
	and is $\sqrt2$ more transparent than its predecessor.
	Thus the visual darkness indicates the probability
	that the process $K_{0∧τ},K_{1∧τ},\dotsc$ passes there.
	
	\def\prdepth{0}
	\newdimen\prvarepsilon\prvarepsilon16pt
	\newdimen\prepsilon\prepsilon512pt
	\advance\prepsilon by-\prvarepsilon
	\newdimen\prbha
	\def\prun#1{
		\edef\prheight{\the\numexpr\prheight-1}
		\ifnum\prheight>0
			\pgfmathsetlength\prbha{#1}
			\draw(\prdepth,\prbha)coordinate(\prdepth)
			\ifnum\prdepth>0
				[line width=2^((\prheight+0)/2)/5,opacity=1/2^((\prdepth-1)/2)]
				(\the\numexpr\prdepth-1)--(\prdepth)
			\fi;
			\ifdim\prvarepsilon<\prbha
				\ifdim\prbha<\prepsilon
					\edef\prdepth{\the\numexpr\prdepth+1}
					{\prun{\prbha/16*\prbha/32}}
					{\prun{(1024-\prbha)/16*\prbha/32}}
				\fi
			\fi
		\fi
	}
\if00
	\def\prheight{16}
\else
	\def\prheight{10}
\fi
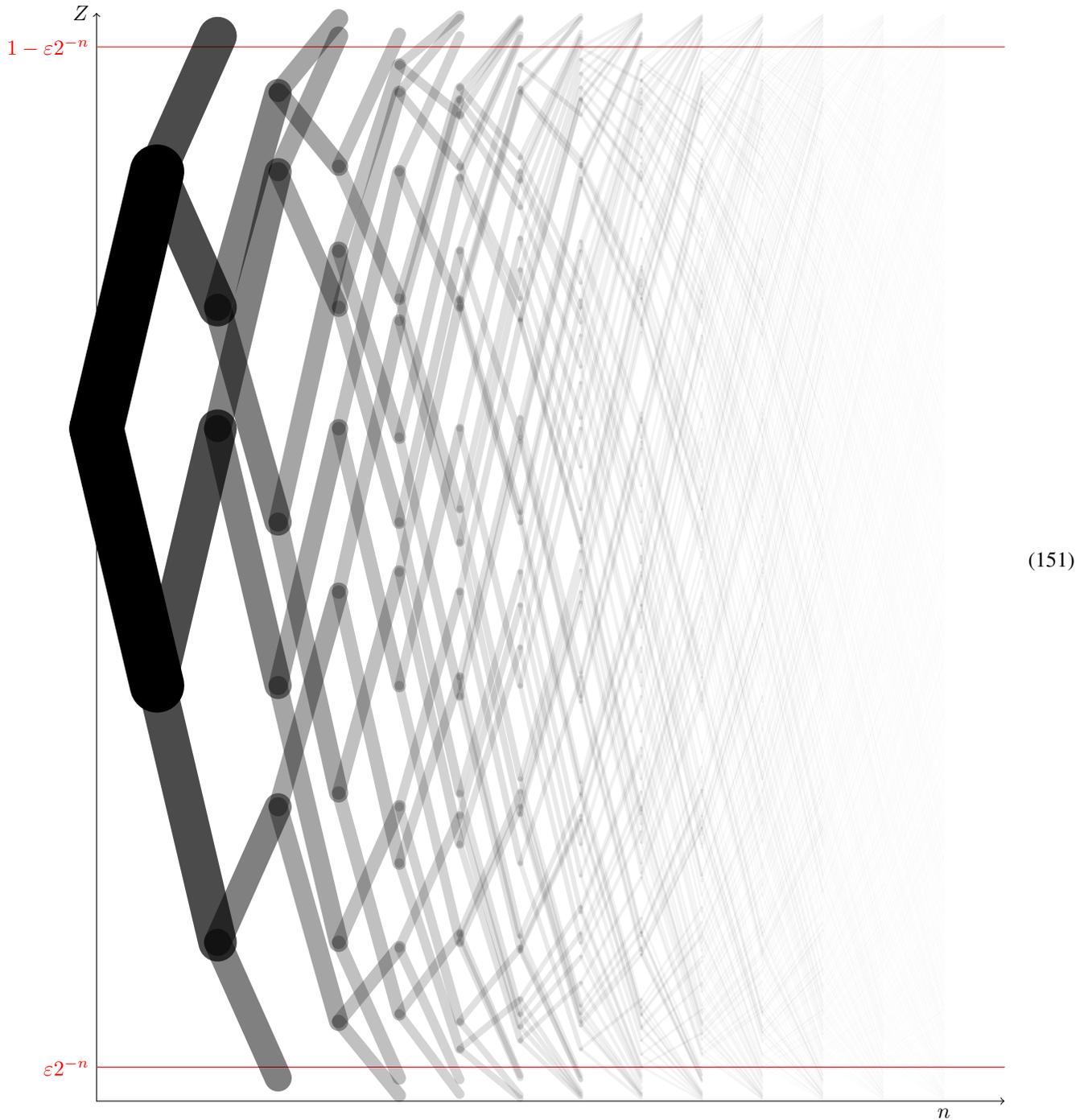
	\[\tikz{
		\draw[->](0,0)--(0,512pt)node[left]{$Z$};
		\draw[->](0,0)--(\prheight-2,0)node[below]{$n$}--+(1,0);
		\draw[red](0, \prepsilon)node[left]{$1-ε2^{-n}$}--+(\prheight-1,0)
		          (0,\prvarepsilon)node[left]{$ε2^{-n}$}--+(\prheight-1,0);
		\prun{.618*512}
	}\]

	Consider this figure \caution{exaggerated} as
	$ε2^{-n}$ shall be smaller than $2^{-n}$,
	about \SI{5}{\micro\meter} on this sheet of paper.
	A 300dpi printer prints in multiple of \SI{85}{\micro\meter}.

\end{document}